\documentclass[twocolumn,superscriptaddress,english,prx,showpacs,longbibliography]{revtex4-2}

\usepackage[colorlinks=true,urlcolor=blue,citecolor=blue,linkcolor=blue]{hyperref}
\usepackage{makecell}
\usepackage{enumitem} 
\usepackage{graphicx}
\usepackage{dcolumn}
\usepackage{bm}
\usepackage{amsmath,amssymb,amsthm,mathrsfs,amsfonts,dsfont}
\usepackage{balance}
\usepackage{color}
\usepackage{array}
\usepackage{tabularx}
\usepackage{longtable}
\usepackage{color}
\usepackage{float}
\usepackage{indentfirst}
\usepackage{txfonts}
\usepackage{booktabs} % professional-quality tables
\usepackage{multirow}

\newcommand{\planar}{\textsc{Planar }}

\usepackage{tocvsec2}
\usepackage{placeins}
\usepackage{verbatim}

\usepackage{physics}
\usepackage{algorithm} 
\usepackage{algorithmic}
\usepackage{subfloat}
\usepackage{graphicx}

\usepackage[all]{hypcap}
\makeatletter
\let\saved@includegraphics\includegraphics
\AtBeginDocument{\let\includegraphics\saved@includegraphics}
\makeatother
\usepackage{braket}
\usepackage{tikz}
\usepackage{float}
\usepackage{amsmath,amssymb,amsthm,mathrsfs,amsfonts,dsfont}
\usepackage{xcolor}

\begin{document}

% 1. Introduction
%     1. introduction of FTQC and current status of QECC and decoding
%     2. Introduction of qecc, stabilizer codes and two ways of decoding
%     3. MWPM and bposd as prominent examples for MWD
%     4. Tensor network decoding as examples for MLD, disadvantages
%     5. introduce planar decoder and its benefits
%     6. paper structure
% 2. Stat-Mech mapping of QEC problems
%     1. QEC to RBIM
%     2. Separate noise approximation and application to MLD
%     3. In some codes, the MLD of uncorrelated noise approximation can be solved in poly-time (surface codes, Honeycomb lattice, etc.)
%     4. how to use Kac-Ward formula to solveuk m'.gvv c the above problem
%     5. Circumstances for periodic boundary condition, pffafian solution
% 3. Decoding planar codes with code capacity noise
%     1. The mapping of planar codes such as surface code & rotated surface code, with separable error model into the calculation of two partition function of planar Ising model.
%     2. threshold analysis
%     3. decoding results
% 4. Decoding repetition code with circuit level noise
%     1. Error models and DEM
%     2. MLD of circuit level decoding
%     3. decoding results
% 5. Discussion
% 6. Supplemental material
%     1. four ways of decoding and their accuracy

\title{Exact Decoding of Repetition Code under Circuit Level Noise}

\author{Hanyan Cao}
\thanks{These authors contributed equally}
\affiliation{CAS Key Laboratory for Theoretical Physics, Institute of Theoretical Physics, Chinese Academy of Sciences, Beijing 100190, China}
\affiliation{School of Physical Sciences, University of Chinese Academy of Sciences, Beijing 100049, China}

\author{Shoukuan Zhao}
\thanks{These authors contributed equally}
\affiliation{Beijing Academy of Quantum Information Sciences, Beijing 100193, China}

\author{Dongyang Feng}
\thanks{These authors contributed equally}
\affiliation{CAS Key Laboratory for Theoretical Physics, Institute of Theoretical Physics, Chinese Academy of Sciences, Beijing 100190, China}
\affiliation{School of Physical Sciences, University of Chinese Academy of Sciences, Beijing 100049, China}
\affiliation{School of Fundamental Physics and Mathematical Sciences, Hangzhou Institute for Advanced Study, UCAS, Hangzhou 310024, China}

\author{Zisong Shen}

\affiliation{CAS Key Laboratory for Theoretical Physics, Institute of Theoretical Physics, Chinese Academy of Sciences, Beijing 100190, China}
\affiliation{School of Physical Sciences, University of Chinese Academy of Sciences, Beijing 100049, China}

\author{Haisheng Yan}
\affiliation{Beijing Academy of Quantum Information Sciences, Beijing 100193, China}

\author{Tang Su}
\affiliation{Beijing Academy of Quantum Information Sciences, Beijing 100193, China}

\author{Weijie Sun}
\affiliation{Beijing Academy of Quantum Information Sciences, Beijing 100193, China}

\author{Huikai Xu}
\email{xuhk@baqis.ac.cn}
\affiliation{Beijing Academy of Quantum Information Sciences, Beijing 100193, China}

\author{Feng Pan}
\email{feng\_pan@sutd.edu.sg}
\affiliation{CAS Key Laboratory for Theoretical Physics, Institute of Theoretical Physics, Chinese Academy of Sciences, Beijing 100190, China}
\affiliation{Centre for Quantum Technologies, National University of Singapore, 117543, Singapore}
\affiliation{Science, Mathematics and Technology Cluster, Singapore University
of Technology and Design, 8 Somapah Road, 487372 Singapore}

\author{Haifeng Yu}
\affiliation{Beijing Academy of Quantum Information Sciences, Beijing 100193, China}
\affiliation{Hefei National Laboratory, Hefei 230088, China}

\author{Pan Zhang}
\email{panzhang@itp.ac.cn}
\affiliation{CAS Key Laboratory for Theoretical Physics, Institute of Theoretical Physics, Chinese Academy of Sciences, Beijing 100190, China}
\affiliation{School of Fundamental Physics and Mathematical Sciences, Hangzhou Institute for Advanced Study, UCAS, Hangzhou 310024, China}

\begin{abstract}
Repetition code forms a fundamental basis for quantum error correction experiments. To date, it stands as the sole code that has achieved large distances and extremely low error rates. Its applications span the spectrum of evaluating hardware limitations, pinpointing hardware defects, and detecting rare events. However, current methods for decoding repetition codes under circuit level noise are suboptimal, leading to inaccurate error correction thresholds and introducing additional errors in event detection. In this work, we establish that repetition code under circuit level noise has an exact solution, and we propose an optimal maximum likelihood decoding algorithm called {\planar}. The algorithm is based on the exact solution of the spin glass partition function on planar graphs and has polynomial computational complexity. Through extensive numerical experiments, we demonstrate that our algorithm uncovers the exact threshold for depolarizing noise and realistic superconductor SI1000 noise. Furthermore, we apply our method to analyze data from recent quantum memory experiments conducted by Google Quantum AI, revealing that part of the error floor was attributed to the decoding algorithm used by Google. Finally, we implemented the repetition code quantum memory on superconducting systems with a 72-qubit quantum chip lacking reset gates, demonstrating that even with an unknown error model, the proposed algorithm achieves a significantly lower logical error rate than the matching-based algorithm.

% As the era of Fault-Tolerant Quantum Computing approaches, the significance of quantum error correction is set to escalate. 
% Although recent times have seen many pivotal advancements in quantum error correction codes, the development of rapid and precise decoders remains elusive. 
% This paper introduces a novel decoding scheme that translates the maximum likelihood decoding (MLD) process of repetition codes with circuit level noise into the computation of a planar Ising spin-glass model partition function. 
% The proposed decoder is the exact MLD in polynomial time of repetition codes under the circuit-level noise. 
% Across all numerical benchmarks on simulations and actual experiments, the Planar decoder demonstrates superior performance compared to the widely-used minimum-weight perfect matching decoder. 

\end{abstract}

\maketitle

Quantum computing has the potential ability to surpass classical computers in tackling challenging practical problems such as crystography~\cite{factor1997,FTneed_2HowFactor20482021}, quantum chemistry~\cite{quantumchemistry2005}, and quantum simulations~\cite{vandamEndEndQuantumSimulation2024}. However, during the present-day noisy intermediate-scale quantum (NISQ) era~\cite{preskillnisq2018quantum}, current quantum computers are not precise enough due to the inherent instability of operations caused by quantum noises. As a critical step in between quantum hardware and quantum algorithms, quantum error correction ~\cite{qec1fowlerintroduction2012,qec2latticesurgery2012,qec3SingleshotError2019,qec4XZZXSurface2021,qec5qccBB2024,qec6_DynamicallyGeneratedLogical2021,qec7_ConstantOverheadFaultTolerantQuantum2023,qec8_landahl83FaulttolerantQuantum2011,qec9_fowler2DColorCode2011,qec10_bombinTopologicalSubsystemCodes2010} aims to correct the errors and decrease the logical error rate to low-enough values that are needed for practical applications. In recent years, there have been numerous experimental advancements in utilizing a variety of quantum error-correcting codes ~\cite{exp2_menendez26ImplementingFaulttolerant2023,reichardt2409demonstration,exp4_hyperproduct2024,exp5_2024,exp6_2024,exp7_color_20211,zhaoRealizationErrorCorrectingSurface2022,bluvstein2023Logical,googlequantumai2023Suppressing,googlenew2024,googlerep}, including surface code~\cite{googlequantumai2023Suppressing,googlenew2024,bluvstein2023Logical,zhaoRealizationErrorCorrectingSurface2022}, color code~\cite{exp7_color_2021}, and subsystem code~\cite{reichardt2409demonstration}. However, due to the limited qubit resources on quantum chips (with typically one hundred qubits) and the limited precision of two-qubit gates (with typically $99.9\%$ fidelity), most experimentally realized quantum codes, e.g. a surface code, only achieved a small distance (up to distance $7$) and high logical error rate (down to $10^{-3}$). 

 Repetition code is the only code that has been realized and demonstrated experimentally to have both large distances and extremely low error rates~\cite{googlerep,leakage2021,leakage2023,midtermrepcode2022}. This is essential for evaluating hardware performance~\cite{midtermrepcode2022, googlerep},  detecting hardware defects like leakage~\cite{leakage2021, leakage2023}, and providing foundations for building scalable fault-tolerant quantum computers. 
The low error rates also pave the way for detecting new physical mechanisms that significantly impact quantum hardware and facilitate the further suppression of quantum errors.
For example, cosmic ray impacts~\cite{cosmicray2024} causing an error floor of $10^{-6}$ have been identified on Google's Sycamore device by using repetition code with distance 25~\cite{googlequantumai2023Suppressing}, and Google's recent experiment with distance $29$ has uncovered previously unobserved error mechanisms at an error floor on the order of $10^{-10}$~\cite{googlenew2024}. 

\begin{figure*}[htb]
    \centering
    \includegraphics[width=\linewidth]{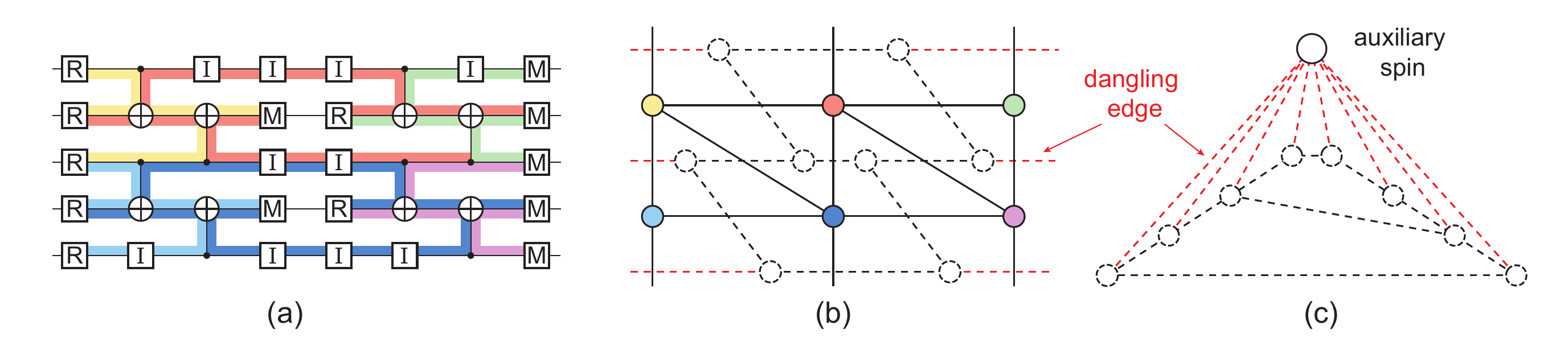}
    \caption{(a) Syndrome measurement circuit and the detecting region of a distance-3 bit-flip repetition code with 2 rounds of measurements. We consider errors caused by the adjacent quantum operations including reset (R), measurement (M), idling (I), and two-qubit gates. Different detecting regions are labeled with different colors. (b) The error graph corresponds to (a). Solid vertices and lines stand for detectors and error mechanisms respectively. Each vertex corresponds to the detecting region with the same color in (a). Dashed vertices and their connections denote the dual graph. The red dashed lines indicate dangling edges that connect to only one detector. (c) The spin glass model used in our exact maximum likelihood decoding algorithm \planar, is mapped from the error graph in (b). An auxiliary spin is added and connected to all dangling edges of the dashed graph in (b).}
    \label{fig:repcir}
\end{figure*}

To achieve an extremely low error rate, the decoding algorithm is essential, as it should be as accurate as possible to avoid introducing additional algorithmic errors.
In recent experiments considering syndrome error (which using the circuit level noise model), decoding was carried out using the minimum-weight perfect matching (MWPM) algorithm and its variants~\cite{dennis2002Topological, higgott2023Sparse, wu2023fusion,correlatedmatching2013,believematching2023}. MWPM algorithm is efficient as it has a polynomial time complexity in the number of qubits. However, as a minimum weight decoder, MWPM algorithms are inherently suboptimal compared to optimal maximum likelihood decoders (MLD). This suboptimality risks introducing additional logical errors during the decoding process.

In this work, we propose an exact maximum likelihood decoding algorithm with polynomial complexity for the repetition code with circuit-level noise. Our algorithm is based on the exact solution to the partition function of spin glasses on planar graphs, so we term our method {{\planar}}. Our algorithm is an optimal decoder for repetition code under circuit level noise, it eliminates the error introduced in the decoding stage with a correct error model. Using our algorithm, we first computed exact thresholds for repetition codes with circuit-level noise under depolarizing and superconducting SI1000 noise models~\cite{mcewen2023Relaxing}, which has not been obtained to our best knowledge. 
Then, we applied our algorithm to Google's repetition code experiments ~\cite{data,googlenew2024}, and show that compared to MWPM used by Google, \planar achieves an average improvement with 25\% in logical error rate, and a higher suppression factor with $\Lambda=8.28$. 
Finally, to evaluate the performance of our algorithm on experimental data with an unknown error model, we implemented a repetition code on superconducting systems using a 72-qubit quantum chip that lacks reset gates. By employing an error model derived through correlation analysis~\cite{googlerep}, our algorithm demonstrated up to an 8.76\% improvement in logical error rates compared to MWPM.

\paragraph*{Repetition code with circuit level noise. ---}

We introduce the repetition code with measurement noise using an illustrative example with distance $d=3$ and $r=2$ rounds of repeated measurements, as shown in Fig.~\ref{fig:repcir}.
%(a), we present the syndrome measurement circuit for a bit-flip repetition code with distance $d=3$. 
The code comprises 3 data qubits and 2 measurement qubits that implement pairwise $Z$ stabilizers. Ancilla qubits are measured and reset for two rounds, and the syndrome measurement results are used to infer whether a logical bit-flip error occurred. All qubits will be measured at the end to obtain the final logical state. %the logical $Z$ logical operator was applied to the initial state of data qubits. 
In this circuit, errors can originate from various sources, including the operation of controlled-not gates, qubit idling (I), reset operations (R), and measurements (M).
%Circuit level noise for a quantum code refers to all possible errors in the syndrome measurement circuit and is often used to distinguish the ideal error model that only considers possible errors on the data qubits~\cite{}.
%Due to the redundancy in the raw measurement results of the syndrome measurement circuit, a 
A standard approach to describing the errors and their effects on logical errors is 
%and managing this redundancy involves the introduction of 
the detector error model (DEM)~\cite{gidney2021Stim,mcewen2023Relaxing}. 
In DEM, a detector is essentially a linear exclusive-or (XOR) combination of multiple measurements and reset results.  
It focuses on detecting discrepancies in consecutive measurements across different circuit layers, offering a more compact representation than the full syndrome measurement results.
As an illustrative example, the syndrome measurement circuits for a bit-flip repetition code and the corresponding detector error model corresponding to the distance-3 repetition code are presented in Fig.~\ref{fig:repcir}(a), where different detecting regions~\cite{mcewen2023Relaxing} are highlighted in different colors. Each detecting region corresponds to a detector associated with a parity check consecutively on $7$ overall qubit measurements and reset. 
Within each region, a specific error that activates the corresponding detector is termed {error mechanism}.
Then we generate an {error graph}~(in Fig.~\ref{fig:repcir}(b)) from the DEM, by mapping $m=6$ detectors and $n=15$ error mechanisms (that are shared by two detecting regions) into vertices and edges respectively. An error mechanism defined on an edge can be treated as the combination of an odd number of Pauli errors that anti-commute with the same set of detecting regions. It will trigger the corresponding detectors defined on the nodes. In this example, there are $2^{15}$ possible error mechanisms grouped into two logical sectors, depending on the commutation relation of the error mechanism and the logical operator.

\paragraph*{Exact decoding.---} 

Given the syndrome measurement $\gamma$, which corresponds to events on detectors in the DEM, decoding is to determine whether a logical operator has been applied. In general, there are two kinds of decoding approaches: minimum weight and maximum likelihood. Minimum weight decoding, exemplified by minimum weight perfect matching, seeks the most likely error consistent with the syndrome measurements. In contrast, the maximum likelihood decoding identifies the most likely logical sector by computing $P(l)$ for each logical sector $l$. $P(l)$ stands for the total probability of all errors that are consistent with the syndrome measures and have the same commutation relations to the logical operator. Maximum likelihood decoding is theoretically optimal, but it is challenging to solve because it belongs to the \#P class in computational complexity. There are no polynomial-time algorithms that can exactly solve this problem in general. In this work, we establish that, for the repetition code with circuit-level noise, exact algorithms exist for solving maximum likelihood decoding with polynomial complexity.

To demonstrate this, we use the example in Fig.~\ref{fig:repcir}. The maximum likelihood decoding computes the probability of all error configurations defined on the edges of Fig.~\ref{fig:repcir}(b) at a specific logical sector $l$ and satisfies the syndrome measurements $\gamma$ on the detectors defined on the nodes. To achieve this, we can sum all Pauli operators $s$ that do not affect the syndrome. 
\begin{equation}\label{eq:pl}
    P(l) = \sum_{s}P[E(s,l,\gamma)],
\end{equation}
where $P[E(s,l,\gamma)]$ is the probability of generating a Pauli error $E(s,l,\gamma)\in \{I,X\}^n$ (as a function of $X$ stabilizer $s$, logical operator $l$, and syndrome $\gamma$) in the error model.
If we consider detectors as $Z$ stabilizers and the error mechanisms it can detect as $X$-type errors, operators $s$ are actually the $X$ stabilizers~\cite{dennis2002Topological} which are represented by dashed nodes in Fig.~\ref{fig:repcir}(b). The nodes and the edges (dashed) that connect them constitute the dual graph of the original detector error graph. 
Given a Pauli error $E$, computing $P(l$) in~Eq.\eqref{eq:pl} can be mapped to the partition function computation of the Ising spin glass defined in Fig.~\ref{fig:repcir}(c). In this figure, each node defines a spin variable $s_i=\pm 1$ denoting whether the corresponding $X$ stabilizer is applied. On the edge $(ij)$ connecting spins $s_i$ and $s_j$ there is a probability $p_{ij}$ that an error mechanism $E_{ij}$ shared by $s_i$ and $s_j$ is generated. This defines a spin-glass coupling~\cite{Chubb_2021}
\begin{equation}
   J_{ij}= \frac{1}{2} \langle E_{ij}, Z_{ij}\rangle \log \frac{1-p_{ij}}{p_{ij}}, 
\end{equation} 
where $E_{ij}$ represents the component of error $E$ acting on the edge $(ij)$. The term $\langle E_{ij}, Z_{ij}\rangle $ is the scalar commutator~\cite{Chubb_2021}, which equals 1 if $E_{ij}$ commutes to the Pauli $Z$ operator acting on the edge $(ij)$, and $-1$ otherwise.

The edges that are only covered by one spin correspond to a dangling edge and map to an external field acting on the spin. The effect of external fields can be fully captured by introducing an auxiliary spin.  Finally and remarkably, this results in that the topology of the corresponding spin glass model with the auxiliary variable is a planar graph, as shown in Fig.~\ref{fig:repcir}(c). Thus, the exact partition function~\cite{PhysRev.65.117} can be computed using a polynomial algorithm such as the Kac-Ward formulism~\cite{kac1952combinatorial, johnson2016learning}. 

\begin{figure*}[tphb]
    \includegraphics[width=0.9\linewidth]{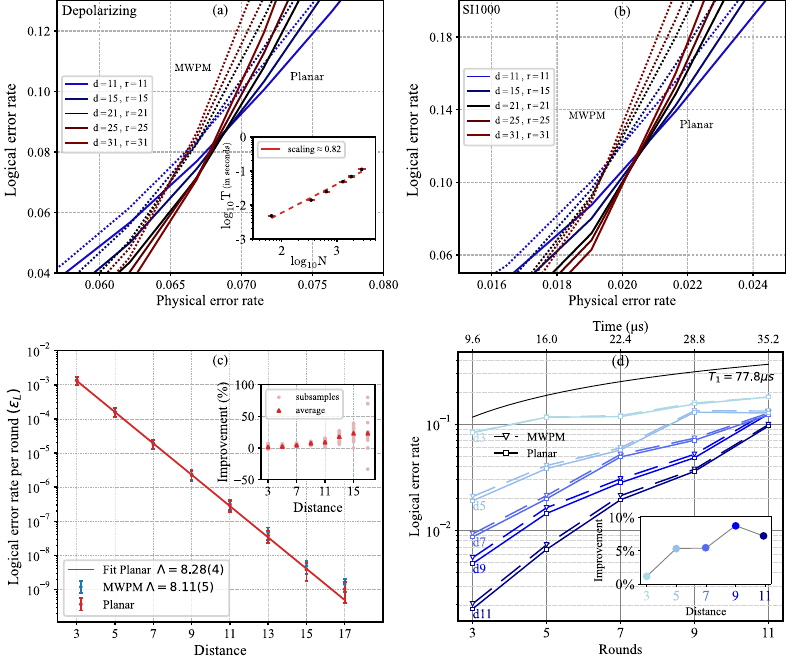}
    \caption{
    Performance of our algorithm in decoding repetition codes under circuit-level noise using various syndrome data.
    (a) Synthetic syndrome data generated using the depolarizing noise model, which is also employed for decoding; (b) Synthetic syndrome data generated by the SI1000 noise model, which is also employed for decoding. In both (a) and (b), we compared the logical error rate obtained by \planar (solid lines) and MWPM (dotted lines). The inset of (a) displays the time complexity of \planar. Each data point represents an average over $10^6$ syndrome shots. Notice that the threshold obtained by \planar is exact for both depolarizing noise and SI1000 noise.
(c) Syndrome data and error model are derived from Google's quantum memory experiments~\cite{googlenew2024}. We compute the logical error rate per round as a function of code distance.  The red solid line represents the linear fitting using data ranging from $d=3$ to 11, as defined in Ref.~\cite{googlenew2024}. We randomly selected $10^6$ samples from Google's datasets of each base $X$($Z$) for $d=[3,5, 7, 9, 11, 13]$ and $2\times10^6$ samples of each base $X$($Z$) for $d=[15, 17]$. The inset displays the improvement of \planar over MWPM. Each data point is averaged over $n_{\text{sub}}$ subsamples with $X$-basis and Z-basis. The light red points represent different subsamples. 
(d) Experimental data are generated on a 72-tunable-coupled-qubits chip, and the error model is derived through the correlation analysis method~\cite{googlerep}. The black solid line represents the guideline for the physical error rate under $T_1 = 77.8\mu\mathrm{s}$. The increasingly deeper shades of blue lines represent distances gradually increasing from 3 to 11. For each distance, solid lines and dashed lines indicate \planar and MWPM respectively. The inset displays the average performance improvement of the \planar method over MWPM. Each data point is averaged over 500,000 samples.
}
\label{fig:2}
\end{figure*}

\paragraph*{Results.---} We evaluate our approach using three examples. In the first example, we use synthetic data generated by depolarizing and SI1000 error models and compute the exact thresholds; in the second example, we revisit the decoding of Google's recent repetition-code quantum memory experiments~\cite{googlenew2024} using their experimental data and error model; in the last example, we conduct the quantum memory experiments using a 72-qubit quantum chip and evaluate the performance of our decoding approach with an error model derived from correlation analysis.

First, we study decoding from synthetic syndromes, where errors were drawn randomly from an error model and syndromes are computed (using Stim~\cite{gidney2021Stim}) for errors and used for decoding. In the case when the error model used in decoding matches the actual error model in generating the data, as a typical inference problem, the decoding is on the Nisimori line~\cite{iba1999nishimori,decelle2011asymptotic} and our exact decoder gives optimal decoding results and the exact threshold. 
In this work, we study two circuit-level error models, the standard depolarizing model and the SI1000 model, which was proposed to model the errors in superconducting devices~\cite{mcewen2023Relaxing}. The results are depicted in Fig.~\ref{fig:2}(a) and (b). We can see that, for both error models with different physical error rates, our planar decoder always gives lower logical error rates than the minimum weight perfect matching. Our data also gives the exact threshold $0.067(7)$ for the depolarizing noise and $0.020(5)$ 
for the SI1000 noise. It is clear that minimum weight perfect matching gives lower thresholds. The inset of Fig.~\ref{fig:2}(a) shows that the time complexity of \planar is $\mathcal O(N^{0.82})$, with $N$ denoting the number of spins and is linear in the number of detectors.
Next, we apply our decoding algorithm to experimental data of quantum memory experiments. We first apply to Google's repetition code experimental data on the \textsc{Willow} quantum chip~\cite{googlenew2024} with $72$ qubits. We randomly chose part of the data from Google's $d=29$ repetition code experiment and used the subsample strategy~\cite{Kelly_2015} to obtain the data of smaller code as well as the error model. We calculated the logical error rate per round~$\epsilon_L = [1-(1-2p_L)^{1/r}]/2$ for various distances and $r=1000$ rounds using this data. 
The comparison of our results with those obtained by Google using MWPM is shown in Fig.~\ref{fig:2}(c). We can see that the improvements of our algorithm to MWPM reach $25\%$ with $d=17$. This reflects that in the original analysis of the error sources~\cite{googlenew2024}, $25\%$ of error should be contributed by the decoding algorithm rather than the unknown sources. We also notice that the improvement increases with code distance because the logical error decays exponentially with distance $d$, with rate defined as $\epsilon_L(d) \approx p_L/r  \propto (1/\Lambda)^{{(d+1)}/2}$. Here $\Lambda$ is an important character of an algorithm and indicates how fast the logical error reduces with $d$ increasing. Here using fitting, we obtained $\Lambda_{Planar} \approx 8.28(4)$ which is significantly improved from $\Lambda_{\mathrm{MWPM}} \approx 8.11(5)$. Note that, we used part of the dataset provided by Google, so we did not get exactly the same $\Lambda$ as in ~\cite{googlenew2024}.

Next, to evaluate the generality and robustness of our decoder under different experimental conditions, we conduct bit-flip repetition-code experiments using a quantum chip with $72$ tunable coupled Transmon qubits with code distances ranging from $3$ to $11$, and up to $11$ rounds of repeated syndrome measurements. In the experiments, we deliberately removed the reset gates and other leakage suppression methods to assess the decoder's effectiveness under a model that deviates from the standard Pauli error model. 
The error model employed in decoding is derived solely through the correlation analysis method~\cite{googlerep}. Like MWPM, we only consider the correlations to align with the error graph shown in Fig.~\ref{fig:repcir}(b) and fail to account for multi-body, long-range correlations that are caused purely by non-Pauli errors.
The results are shown in Fig.~\ref{fig:2}(d). We can see that at all distances and rounds, \planar outperforms MWPM, and the improvement reaches a maximum value of 8.76\% with $d=9$, because \planar captures better the Pauli residual noise.  At $d=11$, the relative improvement decreases abnormally. This reflects that with the system becoming larger, the errors that deviate from simple independent Pauli errors become dominant. In the process of decoding using \planar, we also discovered a qubit in the chip with significantly poor performance. We refer to Supporting Materials for details.

\paragraph*{Discussion.---}
We have established that with a correct error model, the repetition code with circuit-level noise can be optimally decoded using an exact maximum likelihood decoder. We proposed a  polynomial-time decoder -- \planar decoder, utilizing the exact solution of planar spin glasses. Using extensive numerical and physical experiments, we show that under synthetic data with a correct error model, \planar decoder provides the exact threshold for the error model; under experimental data of Google's quantum memory experiments and more noisy data from a 72-qubit experiment without the reset gates, \planar decoder offers lower logical error rates and larger suppression factor $\Lambda$ than the canonical minimum-weight perfect matching based method, it also provides useful information for analyzing hardware errors. The \planar decoder is also very fast, with computational complexity $O(N^{0.82})$. Where $N$ is the number of spins in the spin glass model corresponding to the decoding problem, and is linear in number of detectors. Moreover, our numerical results with different kinds of data, circuits, hardware devices, and errors demonstrated the generality and robustness of our decoder for the practical decoding problem of repetition code experiments.

We note that the \planar method is not limited to decoding repetition codes with circuit-level noise. If the decoding graph of a quantum code exhibits a planar structure under particular error models, the \planar method can be effectively applied. This encompasses both surface and rotated surface codes with independent code capacity noise. Consequently, our approach provides an efficient method for understanding exact thresholds and properties for codes with a planar structure. Moreover, the \planar method can be extended to codes beyond the planar structure. For instance, it can be utilized for decoding codes with the decoding graph possessing a finite genus $g$, with a computational complexity $\mathcal O(4^g)$, yet remains polynomial with the number of spins. We leave this for future study.

\begin{acknowledgments}
This work is supported by Project 12047503, 12325501, 12247104, 12104056, and 92365206 of the National Natural Science Foundation of China, and Innovation Program for Quantum Science and Technology (No.
2021ZD0301802).
F.P. acknowledges the support from the National Research Foundation, Singapore, A*STAR under its CQT Bridging Grant and grants NRF2020-NRF-ISF004-3528. A Python package that is used to complete the numerical experiments of this manuscript can be found in~\cite{planarcode}.
\end{acknowledgments}

\bibliography{merged}

\clearpage

\appendix
\onecolumngrid
\begin{center}
	\textbf{\large Supplementary information for\\ 
``Exact Decoding of Repetition Code with Circuit Level Noise"}
\end{center}

\renewcommand{\theequation}{S\arabic{equation}}
\setcounter{equation}{0} 

\renewcommand{\thefigure}{S\arabic{figure}}
\setcounter{figure}{0} 
\renewcommand{\thetable}{S\arabic{table}}
\setcounter{table}{0} 
\bigskip
\twocolumngrid
%\tableofcontents
% \newpage

\newcommand*{\num}{pi*0.2}

 % define the plot style and the axis style
\tikzset{elegant/.style={smooth,thick,samples=500,cyan}}
\tikzstyle{every node}=[font=\fontsize{10}{15}\selectfont]
\definecolor{green}{RGB}{1,127,1}

\section{Planar decoder For Code Capacity Noise}

\subsection{Statistical Mapping}

Under the Nishimori condition~\cite{Bombin_2012, Chubb_2021}
\begin{equation}
    J({\sigma}) = \frac{1}{4} \sum_{\rho} \langle\sigma, \rho\rangle \cdot \log{p(\rho)} , \quad \sigma, \rho \in \{I, x, y, z\} \;,
    \label{eq:A1}
\end{equation}
decoding of a QECC with $n$ data qubits can be mapped to free energy problem of a random bonds Ising model (RBIM)~\cite{dennis2002Topological}. Where the  $\langle\ ,\ \rangle$ represents the scalar commutator which equals to 1 if two inputs are commute else -1.
In such mapping, stabilizer generators will serve as spins while and data qubits will be (hyper)edges. The Hamiltonian will be formulated as
\begin{equation}
    H_E(s) = - \sum_{\sigma} \sum^n_{i=1} \langle\sigma, E_i\rangle \cdot J(\sigma)_i  \prod_{\{k|\langle\sigma, (g_k)_i\rangle = 1\}} s_k 
    \label{eq:A2}
\end{equation}
where $E$ is a given Pauli error, the subscript $i$ represents the $i$-th qubit. $(g_k)_i$ is the $i$-th component of $k$-th stabilizer. It can easily check the probability of an error $E$ is:
\begin{equation}\nonumber
\begin{split}
    P(E) &= e^{-H_E(s=1)}\\
        &= \exp{\sum^n_{i=1} \sum_{\sigma} \langle\sigma, E_i\rangle \cdot J(\sigma)_i }\\
        &=\exp{\sum^n_{i=1} \sum_{\sigma} \langle\sigma, E_i\rangle \cdot \frac{1}{4} \sum_{\rho} \langle\sigma, \rho\rangle \cdot \log{p(\rho)}}\\
        &=\exp{\sum^n_{i=1} \frac{1}{4}\sum_{\sigma} \sum_{\rho} \langle\sigma, E_i\cdot \rho\rangle \cdot \log{p(\rho)}}\\
        &=\exp{\sum^n_{i=1} \sum_{\rho} \delta_{\rho, E_i} \cdot \log{p(\rho)}}\\
        &=\prod^n_{i=1} p(E_i)
\end{split}
\end{equation}
In the third line, we used the definition Eq.~\ref{eq:A1}. And we used the two properties $\langle A, B\rangle \cdot \langle A, C\rangle = \langle A, BC\rangle$ and $\frac{1}{4} \sum_{\sigma} \langle \sigma, \rho \rangle = \delta_{\rho, I}$ in the next two equations. Additionally, the non-trivial configurations of spins means there are stabilizers acting on the error $E$ and causing a new error $E'$ which belongs to the same equivalence class as $E$. 
Under this mapping, summing all elements which equivalence to $E$ is equal to sum all spin configurations. Therefore the coset probability of a quantum code equals to the partition function of the corresponding Ising model. 
$$ P(L)=\mathcal{Z}_L$$
Calculation of the partition function is a $\#$P hard problem in general. However, there exist polynomial algorithms for special case such as 2D Ising model~\cite{PhysRev.65.117}.
Another difficulty is the $Y$ errors will cause multi-interaction in Hamiltonian. This problem makes the most of quantum codes are not planar, for which the planar Ising model solution is not work.
In order to solve this problem, it is straightforward to decode X-type and Z-type errors independently.
Here we adopt a XZ-separated version of MLD. We consider the X-type and Z-type errors independently i.e. the Y error is the combination of X and Z. The whole code is divided into two parts. A n-qubits Pauli error will also cause two types of syndromes $(s_x, s_z)$. It needs to sum all Z-stabilizers for decoding $s_x$ and sum all X-stabilizers for decoding $s_z$.
\begin{equation}
    \begin{split}
    &L^x = \underset{L_x}{\arg\max} \sum_{S_x\in \mathcal{S}} P(L_x, S_x | s_z) \;,\\
    &L^z = \underset{L_z}{\arg\max} \sum_{S_z\in \mathcal{S}} P(L_z, S_z | s_x)\;, \\ 
    &L = L^x\cdot L^z \;.
    \end{split}
    \label{eq:A3}
\end{equation}
    
% \end{gather}    

In fact the Eq.~\ref{eq:A3} is an approximation of Eq.~\ref{eq:A1} for general error model. And its accuracy is depend on the correlation between X and Z. Empirically, most of minimum weight algorithms are finding the minimum weight error of X and Z separately, like Minimum Weight Perfect Matching (MWPM)~\cite{higgott2021pymatching, higgott2023Sparse} and belief propagation augmented by ordered statistics decoding (BPOSD) algorithm ~\cite{Panteleev2021degeneratequantum, Roffe2020}. Therefore, XZ-separation version of MLD will be better than MWPM theoretically. An approximately scheme that has been used to compute coset probabilities, utilize the XZ-separated MLD is renormalization group (RG)~\cite{ducloscianci2010renormalization}. In this work we proposed that the exact coset probabilities can be calculated in polynomial time for some special codes. At first this method considers the XZ-separated MLD. Then it converts the code into a planar Ising model and calculates the partition function using the strict algorithm of planar Ising model.  

\begin{figure*}[!htbp]
    \centering
    \includegraphics[width=0.8\linewidth]{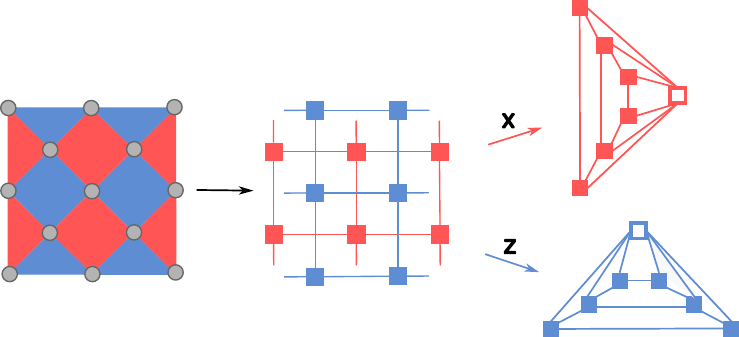}
    \caption{Illustration of separating the distance 3 surface code. In the left graph, the red(blue) blocks are X(Z)-stabilizers and the gray nodes denote qubits which are (hyper)edges in RBIM. The red(blue)-subgraph at middle represents X(Z)-Ising model. The squares denote X(Z)-spins respectively. In the right graph, we add an additional spin for X(Z)-Ising model to construct a planar Ising model without external fields.}
    \label{fig:sur}
\end{figure*}

There exists discussion of a special case code can be solved rigorously, the surface code~\cite{bravyi2014Efficient}. However our work gives more general results. If the X or Z subgraph of a correction code corresponds to a planar graph, with XZ-uncorrelated noise, their MLD procedures can be transformed into partition function calculations of planar RBIM. And it can be solved in polynomial time in their system sizes. It is worth noting that it is only necessary that either the X or Z subgraph be planar, since X and Z are dual to each other, and when one is planar, the other is naturally planar. Surface codes, for example, have four-body interactions that vanish under uncorrelated noise as Fig.~\ref{fig:sur} show. Each edge connects to two spins and the graph is obviously planar. And there exist polynomial time algorithms for the calculation of exact free energy of the planar Ising model~\cite{kac1952combinatorial}. Therefore, it can decode Surface code with uncorrelated noise exactly in polynomial time. Further more it can also use the uncorrelated approximation to decoding more complex error model such as depolarizing model. 

\subsection{Kac-Ward solution}
We adopt the Kac-Ward determinant formula to compute the partition function of planar Ising model~\cite{kac1952combinatorial, johnson2016learning}. The main conclusion of this formula is that the partition function of a zero-field Ising model defined on a planar graph $G(E,V)$ is proportional to the determinant of a matrix. 
\begin{equation}
    Z = 2^n \left(\prod_
    {(i,j)\in E} \cosh{\theta_{ij}}\right)\det(I-W)^{\frac{1}{2}} \;,
\end{equation}
where the $\theta$ satisfies 
the definition $\tanh{\theta_{ij}} \triangleq J_{ij}$, and the $W=AD$, $D$ is a $2|E|\times2|E|$ diagonal matrix $D_{ij,ij} = J_{ij}$. The definition of  $A$ is under follow:
\begin{equation}
  A_{ij, kl} =
    \begin{cases}
      \exp(\frac{i}{2}\phi_{ijl}), &j=k \ and \  i\neq l\\
      0, & otherwise
    \end{cases}\;.
\end{equation}

The $\phi_{ijl}$ is clock-wise angle difference between directed edges $(ij)$ and $(jl)$. The determinant can be computed by Gaussian elimination with the $O(n^3)$ complexity. And the complexity of this calculation will reduce to $O(n^{\frac{3}{2}})$, if the sparse structure of the matrix is used.

\subsection{Numerical Results}

\begin{figure*}[!htbp]
\center
\includegraphics[scale=0.47]{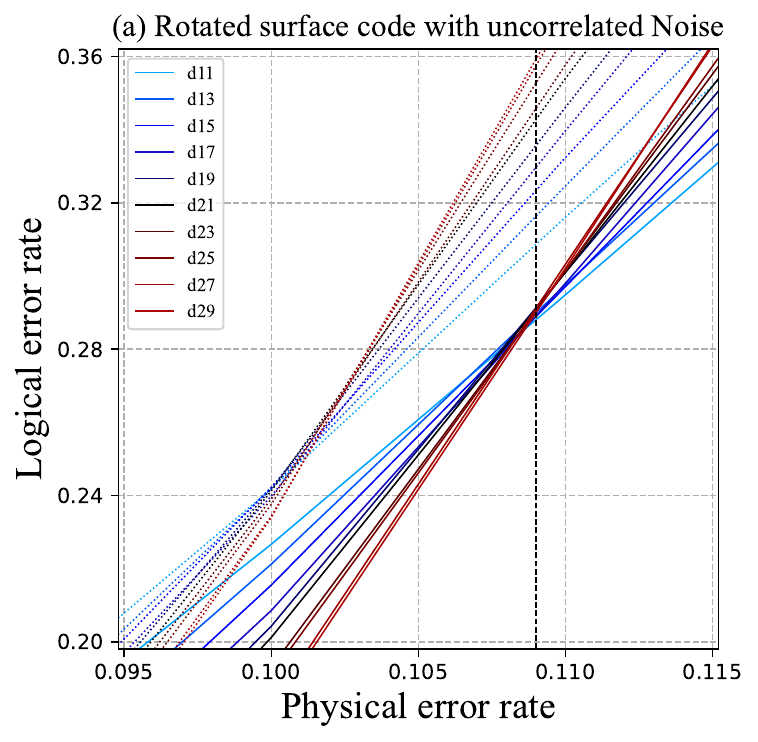}\hspace{0pt}
\includegraphics[scale=0.47]{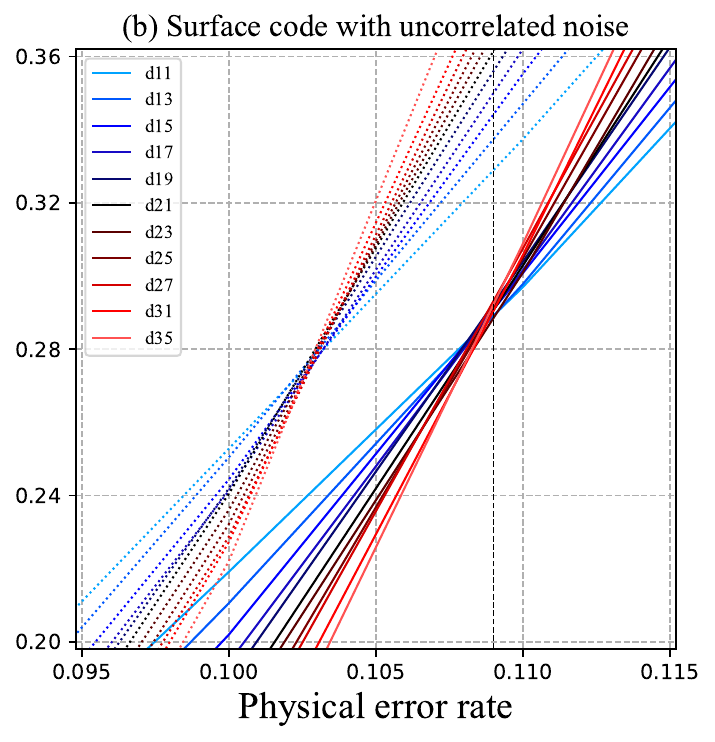}\hspace{0pt}
\includegraphics[scale=0.47]{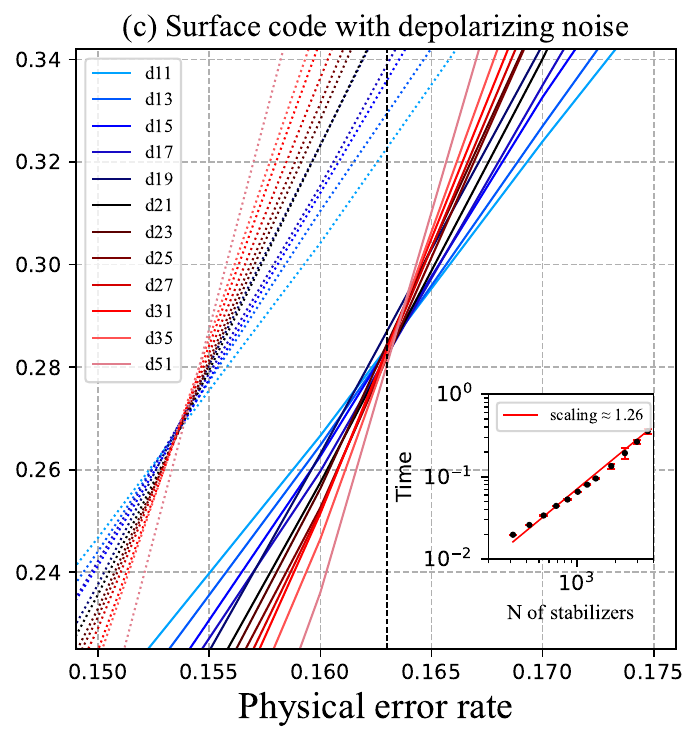}\hspace{0pt}
\caption{Simulation on (Rotated) Surface codes with code capacity noise. We compared logical error rate calculated by Planar (solid lines) and MWPM (dashed linse). Each data point has $10^5$ shots.} 
\label{fig:res}
\end{figure*}

The first code family we considered is the Surface code. As previous discussion that the Surface code is equivalent to a Ising model with (three)four bodies interactions which is due to the existence of Y error on a single qubit. Under the uncorrelated noise, 
\begin{equation}
    \begin{split}
    &p(I) = (1-p_x)(1-p_z),\\
    &p(x) = p_x(1-p_z),\\
    &p(z) = (1-p_x)p_z,\\
    &p(y) = p_xp_z
    \end{split}
    \label{eq:A4}
\end{equation}
The correlations between X and Z types of stabilizers disappear i.e. $J(Y) = 0$ in Eq.~\ref{eq:A2}. The Hamiltonian can be reduce to following term
\begin{equation}
\begin{split}
H_E(s) = - \sum^n_{i=1} (-1)^{\delta_{E_i, z}}\ J(x)_i  \prod_{\{k| (g_k)_i = z\}} s_k\\
- \sum^n_{i=1} (-1)^{\delta_{E_i, x}}\ J(z)_i  \prod_{\{k| (g_k)_i = x\}} s_k
\end{split}
\label{eq:A5}
\end{equation}

From above equation, one can find the X and Z subsystem are independent. The partition function of the whole system is product of partition functions of two subsystem $\mathcal{Z} = \mathcal{Z}_x \cdot \mathcal{Z}_z$. And the $L_x$ and $L_z$ are independent consequently $P(L_x, L_z) = P(L_x)\cdot P(L_z)$. At this point, The graph of original ising model can be separated to X and Z subgraphs. Each subgraph is a $d\times (d-1)$ lattice. And spins at top and bottom(left and right) interact with the external fields. These external fields arise because qubits on the surface code boundary are only subject to a single X or Z check. This is shown as Fig.~\ref{fig:sur}.(middle). 

The presence of external fields causes the strict solution of the Ising model fail. However this problem can be handled by introducing an additional spin. Each boundary spins interact with this new spin and the strength of coupling is equal to $J(\sigma)_{\partial}$, the $\partial$ denotes a boundary qubit of surface code as Fig.~\ref{fig:sur}. (right). Obviously the new graph is also planar. The partition function of the new Ising model $Z_a = 2 Z$. This is because the introduction of the new spin gives the system $z_2$ symmetry. From the perspective of group theory, this additional spin represents a new stabilizer that can be obtained as a product of the original stabilizers. Therefore, it is still an element of the original stabilizer group. According to the rearrangement theorem, when summing over all generators, considering this element is equivalent to summing over all group elements twice.

Error correction will be achieved by calculating four times the partition functions. The syndrome and logical inputs are changed by changing E in Eq.~\ref{eq:A5}. And they will not change the structure of graph. This error correction process is strictly MLD for uncorrelated noise, because we calculate the free energy exactly. At the same time it is a good approximation for correlated noise. 
We test \textit{Planar} on uncorrelated error model. It will gives the exact result of MLD at uncorrelated case. The threshold is about $p_{uc}\approx0.109$, which can be read out from the Fig.~\ref{fig:res}(middle) approximately. This result is consistent with the analysis from a statistical physics perspective ~\cite{Chubb_2021}.
Numerical decoding are also simulated on depolarizing error model. The prior probability $p_x = p_z = 2\epsilon/3$, where $\epsilon$ is physical error rate of depolarizing model. This modification thinks of the Y error as being composed of independent X and Z simply and is always used in MWPM~\cite{higgott2021pymatching} and BPOSD~\cite{Roffe2020}. One can find the point where the curves intersect is between 0.162 and 0.164. This result is consistent with the theoretical value $p=\frac{3}{2}p_{uc}\approx
0.164$~\cite{Bombin_2012}. And the time scaling of our algorithm is about 1.26.

Another code family, we test the \textit{Planar} is the rotated surface code (RSC)~\cite{Bombin2007}. It is just cutting and rotation of the Surface code lattice and has almost the same threshold as the Surface code. However it has higher code rate, and used in a wide variety of experiments~\cite{googlequantumai2023Suppressing}. Same as the analysis of Surface code. The X(Z)-subgraph of RSC is also planar. And all open edges need to connect to a new spin. The difference is there exist boundary spins interacting with two external fields. A simple way to deal with these two dangling edges is to merge them together. And the weight of new edge is sum of weights of two original edges. 
And the numerical results are shown in Fig.~\ref{fig:res}.

\section{Circuit Level Noise and Detector Error Model}

In order to better configure the parameters of the decoder, we need to closely align the error model with real QEC experiments. This requires us to consider in detail all possible error locations and error types that can occur when compiling a quantum circuit for a QEC code. We assume that the error locations accompanies all effective quantum operations (including idling), such as quantum gates and measurements. Due to the nature of stabilizer codes, we consider all possible errors to be Pauli errors. These errors and their corresponding probabilities form an error model. The most commonly used and well-known model is depolarizing error model, assumes that Pauli $X$, $Y$, or $Z$ errors occur at any possible error location with the same probability. Another model used in this paper is the newly developed SI1000~\cite{mcewen2023Relaxing}, which is an empirical model derived from the statistical behavior of superconducting quantum computing equipment. Both models use a single parameter $p$ to configure the Pauli errors across all possible locations and types. The difference is that SI1000 assigns different weights to error rates at different locations to better simulate the performance of superconducting devices. The specific configuration of the error model used in this paper is detailed in~\cite{mcewen2023Relaxing} Table~D.2.

\begin{figure}
    \centering
    \includegraphics[width=0.8\linewidth]{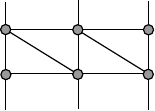}
    \caption{DEM graph, each gray node represents a detector, while each edge corresponds to an error mechanism}
    \label{fig:simple_dem}
\end{figure}

After fixing the compiled quantum circuit, and considering the circuit-level noise error model, certain errors are equivalent to each other (also known as gauge equivalent~\cite{gauge}) which means they would produce the same syndrome, i.e., resulting in identical detectors and having the same effect on the encoded logical qubit. Therefore, we can bundle these equivalent errors together and group them, with each group referred to as an "error mechanism." Similar to the circuit-level noise error model, each error mechanism can also be assigned a probability value. This probability represents the likelihood of any odd number of errors occurring within the group corresponding to that error mechanism. This bundled model, composed of error mechanisms and detectors, is referred to as the detector error model (DEM). As shown in Fig.~\ref{fig:simple_dem}, each edge represents an error mechanism with a weight equal to its probability value. The nodes it connects are the corresponding detectors, each of which can be in one of two states: flipped or not flipped.

In the experimental process, the correspondence between error mechanisms and detectors remains the same as in the simulation results once the circuit is finalized. However, the probability value associated with each error mechanism is no longer derived from the combination of multiple corresponding Pauli errors. Instead, it is determined through correlation analysis~\cite{googlerep}, which directly uses multiple experimental samples to establish the relationships between detectors. This correlation is used in place of probability values to better approximate the error probability distribution in the actual circuit, thereby achieving more optimal decoding performance.

\section{Subsample Strategy}

\begin{figure*}[htbp!]
    \centering
    \includegraphics[width=\linewidth]{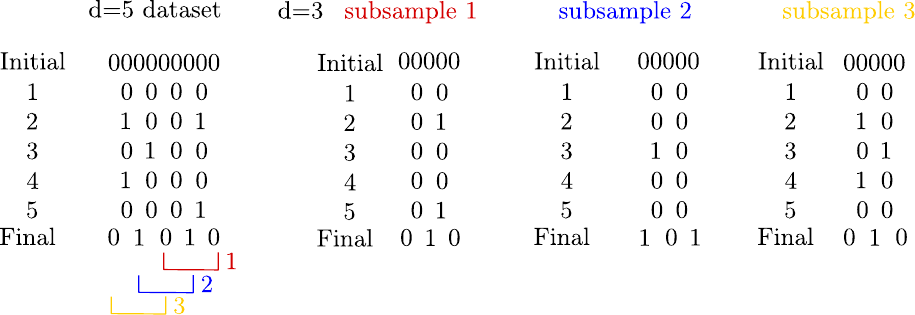}
    
    \caption{Example of deriving subsamples of d=3 code from d=5 dataset.}
    \label{fig:subsample}
\end{figure*}

In this section, we provide a detailed explanation of the subsampling strategy. This approach involves deriving samples from a lower-distance code by extracting them from a higher-distance code. Figure \ref{fig:subsample} illustrates an example where d=3 samples are obtained from a d=5 code.
For a d=5 repetition code, there are 9 initial measurement results (default in the Z-basis), corresponding to 9 qubits, of which 5 are data qubits and 4 are ancilla qubits. During each round of syndrome measurement, the 4 ancilla qubits are measured to yield syndrome results. In the final round, all 5 data qubits are measured. From this setup, we can derive 3 subsamples corresponding to a d=3 repetition code.
Subsampling offers several practical advantages. First, it reduces the number of trials needed to acquire experimental data. Without subsampling, experiments would have to be conducted separately for each distance code. However, by employing subsampling, it suffices to perform experiments on the code with the highest distance and then extract data for lower-distance codes from the same dataset. Additionally, subsampling ensures that the experimental data for different distances are obtained under identical conditions, using the same equipment. This consistency enhances the reliability and comparability of the data across various distances.

\begin{figure}[htbp!]
    \centering
    \includegraphics[width=0.8\linewidth]{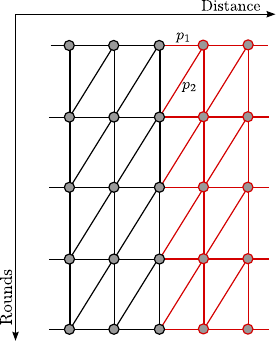}
    
    \caption{Sub-DEM of d=3 in DEM of d=5 repetition code. The red nodes and lines corresponding to the DEM of subsample 1 in Fig.\ref{fig:subsample}.}
    \label{fig:subdem}
\end{figure}

Unlike MWPM, decoding subsampled data using \textit{Planar} requires additional processing of the DEM of the higher-distance code. In Figure \ref{fig:subdem}, we illustrate the Sub-DEM for the d=3 subsample 1 repetition code derived from the DEM of a d=5 repetition code. Compared to Figure \ref{fig:simple_dem}, the red region in Figure \ref{fig:subdem} contains several additional beveled edges. For MWPM, these beveled edges can be directly handled and do not introduce any additional complexity. However, when using \textit{Planar}, it is necessary to combine the probabilities associated with these beveled edges with the probabilities of the transverse edges connected to the same node. The combined probability is calculated as $p = p_1(1-p_2) + p_2(1-p_1)$. After this calculation, the beveled edges are removed to streamline the decoding process.

\section{Experiment Details}

\subsection{Fabrication}

\begin{figure}
    \centering
    \includegraphics[width=0.99\linewidth]{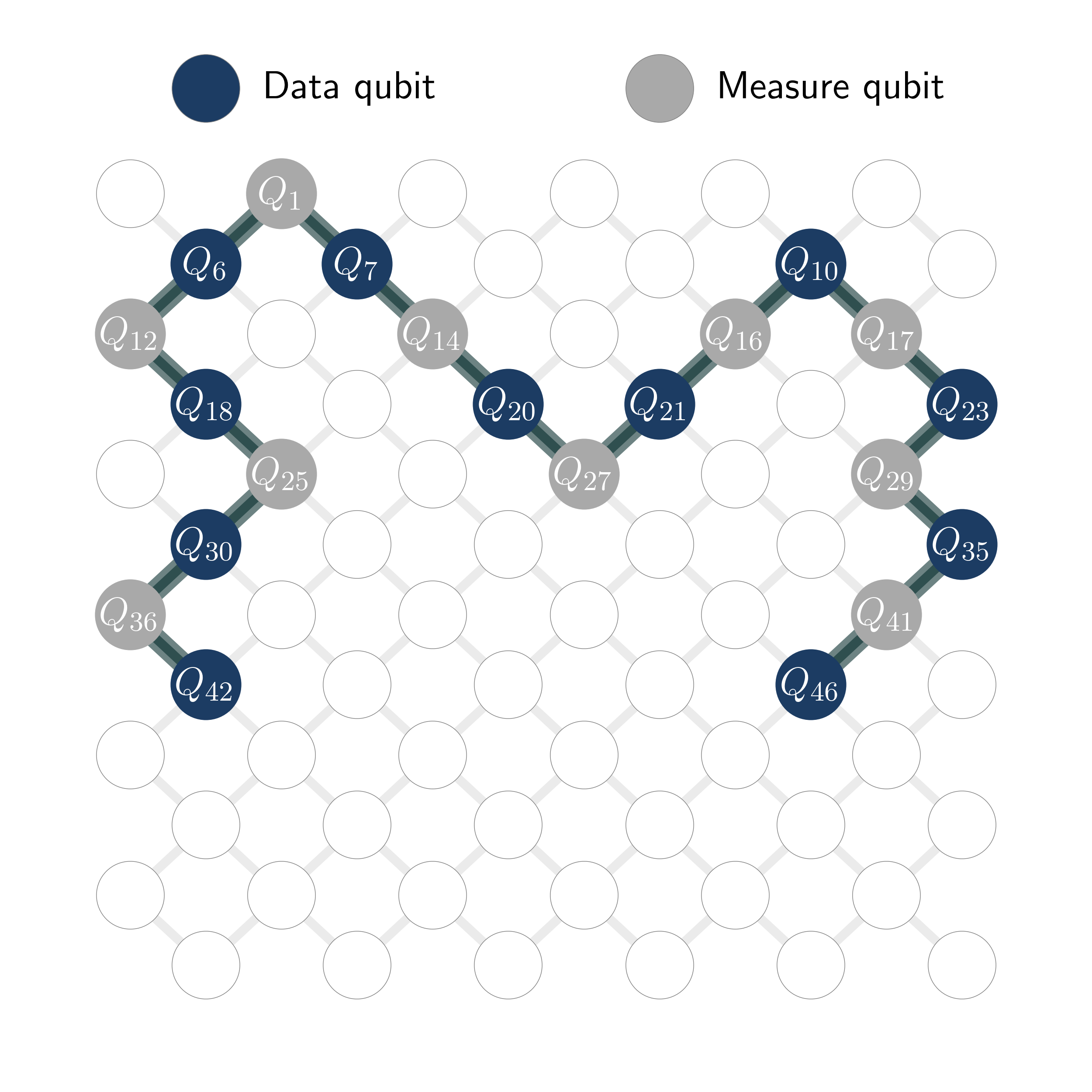}
    \caption{Layout of active qubits in the 72-tunable-coupled-qubit chip. gray circles represent measurement qubits, while dark blue circles indicate data qubits used in the distance-11 repetition code. In the experiment, codes of different distances utilize overlapping sets of qubits. The labels inside the circles denote the unique identifier of the qubits.}
    \label{fig:layout}
\end{figure}

The superconducting quantum processor used in this experiment consists of 72 qubits and 121 transmon-like couplers. As illustrated in Fig.~\ref{fig:layout}, qubits are represented as circles, while the lines connecting the circles indicate couplers. The processor is fabricated using a flip-chip architecture~\cite{Li2024}, with the carrier chip measuring 38 mm × 45 mm and the top chip measuring 18 mm × 20 mm. The circuits on the top chip are fabricated on a c-plane sapphire wafer, while those on the carrier chip are fabricated on a silicon wafer.

The fabrication process begins with annealing the sapphire wafer in a chamber annealing furnace at 1200 ℃ for 3 hours in an atmospheric environment. The silicon wafer is pre-treated for 2 minutes using an HF solution. A 200-nm-thick tantalum film is then deposited on both wafers via a DC magnetron sputtering system. The base structures, including transmission lines and flux lines on the carrier chip and resonators and qubit capacitors on the top chip, are patterned using photo-lithography with a laser direct-writing lithography (DWL) system. Following development, reactive ion etching (RIE) with CF4 gas is performed to remove unwanted metallic regions, finalizing the base structures for subsequent processing steps.

For the top chip, after the base circuits are defined, the chip undergoes a piranha solution bath (2:1 $\mathrm{H}_2\mathrm{SO}_4$:$\mathrm{H}_2\mathrm{O}_2$) for 20 minutes, followed by an ultrasonic bath and a 10-minute rinse in deionized (DI) water. This piranha treatment is effective in improving the surface quality of the tantalum films. Next, Dolan-bridge $\mathrm{Al}/\mathrm{AlO}_x/\mathrm{Al}$ Josephson junctions are patterned using a double-angle shadow evaporation process in a 4-chamber e-beam evaporation system. After a lift-off process, the top chip is prepared for packaging.

For the carrier chip, aluminum air-bridges are fabricated on top of the base circuits. The bridge-shaped profile is created using a reflow method after the S1813 photoresist is patterned and developed with MF319 developer. Subsequently, a 500-nm-thick aluminum layer is deposited by evaporation, followed by lift-off steps. Indium bump structures are then patterned using a 2-layer resist photolithography process, and a 20-$\mu$m-thick indium film is deposited via thermal evaporation. The lift-off of indium is achieved by immersing the chip in an 80 ℃ NMP bath for over 8 hours.

Once both chips are ready, flip-chip bonding is performed using an SET Accura system. Accurate alignment of the two chips is ensured by a coaxial optical microscope, and a pressure of 6 kg is applied to achieve a uniform gap of 20 $\mu$m between the chips. This alignment and gap uniformity are subsequently verified using a scanning electron microscope (SEM).

\subsection{Measurement setup}

\begin{figure*}[htbp]
    \centering
    \includegraphics[width=\linewidth]{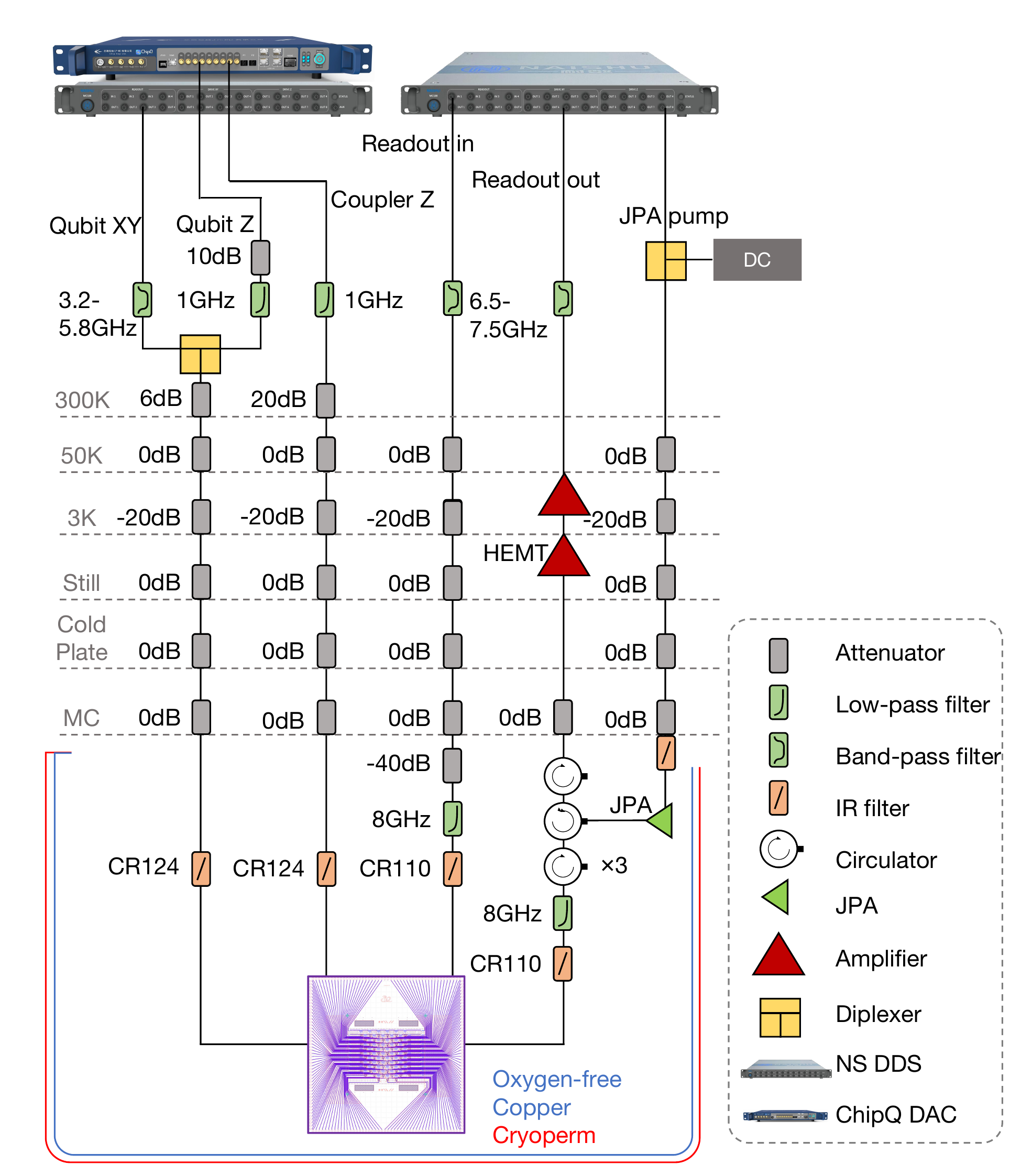}
    \caption{ A schematic of the measurement system, including cryogenic and room temperature wiring, various microwave devices and control system, amplifying circuit,
measurement electronics, and electromagnetic shielding.}
    \label{fig:setup}
\end{figure*}

Our experimental setup shown in Fig.~\ref{fig:setup} is identical to that described in Ref.~\cite{zhao_quantum_2024}. The room-temperature electronics include systems for qubit readout, fast Z control, XY control, and Josephson parametric amplifier (JPA) control. For microwave generation and mid-circuit measurements, we utilize RF Direct Digital Frequency Synthesis (DDS) with a measurement repeat period of 3.2 $\mu$s.

For fast Z control, the equipment is equipped with hardware-implemented 6th-order infinite impulse response (IIR) filters on each channel, enabling real-time compensation for Z distortions. To further mitigate the effects of noise and harmonic distortions, a bandpass filter and an infrared (IR) filter are incorporated into the circuit design.

\subsection{Optimization of readout}

\subsubsection{Method to state classification}

Experimentally, the IQ signal of a qubit’s readout resonator can be used to represent the qubit’s state. To classify the states $\ket{0}$ and $\ket{1}$ of a qubit, the qubit should be initialized in $\ket{0}$ and $\ket{1}$ using the $I$ gate and $X$ gate, respectively. The IQ signals for these states are then measured multiple times and labeled as $S_0$ and $S_1$. The data points from $S_0$ and $S_1$ can be plotted on the complex plane, where a boundary that optimally separates the two sets is determined. This boundary delineates the regions on the complex plane corresponding to the distributions of the two states.

% \begin{figure}[H]
%     \centering
%     \begin{tikzpicture}
%         \node[anchor=south west,inner sep=0] (image) at (-1.5,0) {\includegraphics[width=0.4\textwidth]{figures/sup/test.png}};
%         \draw[blue,dotted] (5.3-0.6,4.5-0.8)--(5.8-0.6,4.5-0.8);
%         \node[color=blue] (B) at (6-0.6,4.5-0.8) {$\ket{0}$};
        
%         \draw[red,dotted] (5.3-0.6,4-0.8)--(5.8-0.6,4-0.8);
%         \node[color=red] (B) at (6-0.6,4-0.8) {$\ket{1}$};
        
%     \begin{scope}[x={(image.south east)},y={(image.north west)}]
%     \end{scope}
    
%     \end{tikzpicture}
%     \caption{\textbf{IQ signal of experiment}. The blue and yellow points are experimental IQ signal of state $\ket{0}$ and $\ket{1}$, respectively. The dotted blue and red circles are experimentally corresponding to the circles $\bar{S}_0$ and  $\bar{S}_1$, respectively. The upper histograms represent the statistical distributions of the two states along the direction of $\bar{S}_0$ and $\bar{S}_1$. }
%     \label{shots}
% \end{figure}

\begin{figure*}[htbp]
    \centering
    \includegraphics[width=0.9\linewidth]{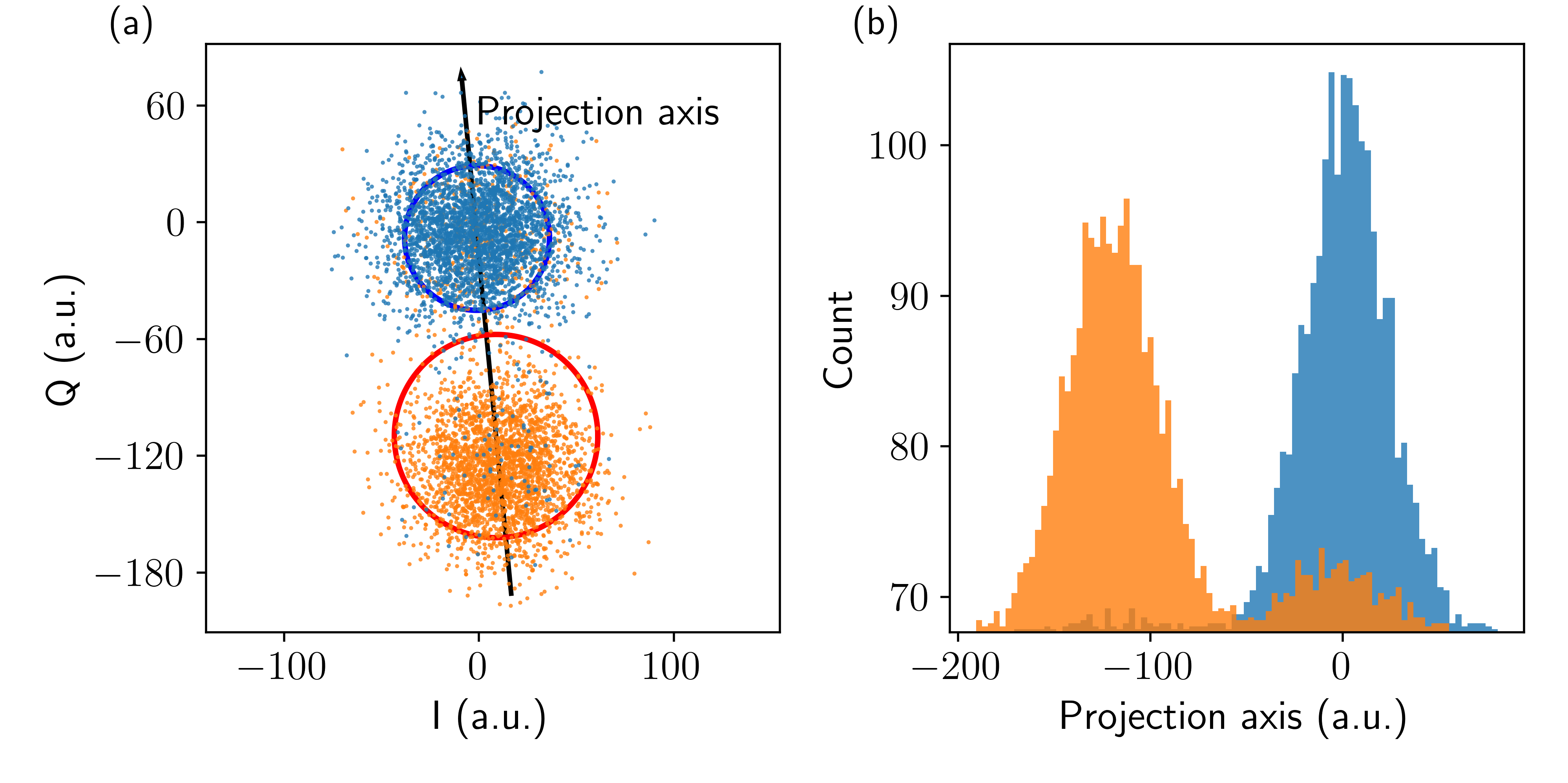}
    \caption{\textbf{IQ signal of experiment}. The blue and yellow points are experimental IQ signal of state $\ket{0}$ and $\ket{1}$, respectively. The dotted blue and red circles are experimentally corresponding to the circles $\bar{S}_0$ and  $\bar{S}_1$, respectively. The upper histograms represent the statistical distributions of the two states along the direction of $\bar{S}_0$ and $\bar{S}_1$. }
    \label{fig:singleshot}
\end{figure*}

\begin{figure}[H]
    \centering
    \begin{tikzpicture}
        \node at (-1,1.3)[below]{\text{(a)}};
        \node at (-1,1.3-3.5)[below]{\text{(b)}};
        \draw[dotted] (-1,0)--(7,0);
        
        \node at (1.35,-0.25)[right]{$\bar{S}_0$};
        \draw[dotted,blue] (1.5,0) circle (1); % circle S_0
        \fill (1.5,0) circle (1.5pt);
        \draw[thick,-latex] (1.5,0)--node[above,sloped] {$\sigma_0$}(1.5+0.707,0.707);
        
        \node at (2.5,0)[above]{${P}_0$};
        \fill (2.5,0) circle (1.5pt);
        \draw[blue] (1.2,0) circle (1.3); % circle S_0'
        \fill (1.2,0) circle (1.5pt);
        \node at (1.35,-0.25)[left]{$\bar{S}_0^{'}$};
        \draw[thick,-latex] (1.2,0)--node[above,sloped] {$\sigma_0^{'}$}(1.2-1.1258,0.65);

        \draw[dotted,red] (4.5,0) circle (1.25); % circle S_1
        \fill (4.5,0) circle (1.5pt);
        \node at (4.65,-0.25)[left]{$\bar{S}_1$};
        \draw[thick,-latex] (4.5,0)--node[above,sloped] {$\sigma_1$}(4.5-0.707*1.25,0.707*1.25);
        
        \node at (3.25,0)[above]{${P}_1$};
        \fill (3.25,0) circle (1.5pt);
        \draw[red] (4.7,0) circle (1.45); % circle S_1'
        \fill (4.7,0) circle (1.5pt);
        \draw[thick,-latex] (4.7,0)--node[above,sloped] {$\sigma_1^{'}$}(4.7+0.866*1.45,0.5*1.45);
        \node at (4.6,-0.25)[right]{$\bar{S}_1^{'}$};

        \draw[dotted] (-1,-3.5)--(7,-3.5);
        
        \node at (1.5+0.5,-3.5)[below]{$\bar{S}_0$};
        \fill (1.5+0.5,-3.5) circle (1.5pt);
        \draw[dotted,blue] (1.5+0.5,-3.5) circle (1); % circle S_0, down
        \node at (2.5+0.5,-3.5)[above]{${P}_0$};
        \fill (2.5+0.5,-3.5) circle (1.5pt);
        \draw[thick,-latex] (2,-3.5)--node[above,sloped] {$\sigma_0$}(2-0.707*1,-3.5+0.707*1);
        
        \draw[blue] (0.8+0.5,-3.5) circle (1.25); % circle S_0', down
        \draw[thick,-latex] (1.3,-3.5)--node[above,sloped] {$\sigma_0^{'}$}(1.3-0.866*1.25,-3.5+0.5*1.25);
        \fill (1.3,-3.5) circle (1.5pt);
        \node at (1.3,-3.5)[below]{$\bar{S}_0^{'}$};

        \draw[dotted,red] (3.3+0.5,-3.5) circle (1.25); % circle S_1, down
        \fill (3.3+0.5,-3.5) circle (1.5pt);
        \node at (3.3+0.5,-3.5)[below]{$\bar{S}_1$};
        \node at (2.05+0.5,-3.5)[above]{${P}_1$};
        \fill (2.05+0.5,-3.5) circle (1.5pt);
        \draw[thick,-latex] (3.8,-3.5)--node[above,sloped] {$\sigma_1$}(3.8+0.707*1.25,-3.5+0.707*1.25);
        
        \draw[red] (3.95+0.5,-3.5) circle (1.45);  % circle S_1', down
        \fill (3.95+0.5,-3.5) circle (1.5pt);
        \node at (3.95+0.5,-3.5)[below]{$\bar{S}_1^{'}$};
        \draw[thick,-latex] (4.45,-3.5)--node[above,sloped] {$\sigma_1^{'}$}(4.45+0.866*1.45,-3.5+0.5*1.45);
        
    \end{tikzpicture} 
    \caption{
        \textbf{State classification circles of qubits in our experiment}. 
        $\bar{S}_i$ and $\sigma_i$ are expectation and standard deviation of $S_i$ (i=0,1). 
        $\bar{S}_i^{'}$ and $\sigma_i^{'}$ are center and radius of state $\ket{i}$ classification circle of qubit. Circle $\bar{S}_i^{'}$ is shown in \textbf{a} for $\left\lvert \bar{S}_0 - \bar{S}_1 \right\rvert \geq \sigma_0 + \sigma_1$, while in \textbf{b} for $\left\lvert \bar{S}_0 - \bar{S}_1 \right\rvert < \sigma_0 + \sigma_1$. 
        }
    \label{CC}
\end{figure}

To improve readout fidelity and sampling efficiency during bitstring measurement, we classify the $\ket{0}$ and $\ket{1}$ states of a qubit using two regions defined by circles on the complex plane. The method for determining the parameters of these circles (shown as solid circles in Fig.~\ref{CC}) is as follows. Let $\bar{S}_i$ and $\sigma_i$ denote the expectation and standard deviation of $S_i$ ($i=0,1$), where $\bar{S}_i$ is a complex value. As illustrated in Fig.~\ref{CC}, the dotted circle centered at $\bar{S}_i$ has a radius $\sigma_i$, while the solid circle $\bar{S}_i^{’}$, representing the classification boundary for state $\ket{i}$, has a larger radius $\sigma_i^{’}$ ($>\sigma_i$). The experimental IQ signals and their respective expectations $\bar{S}_i$ are shown in Fig.~\ref{fig:singleshot}.

The point $P_i$ is defined as the intersection of the dotted circle $\bar{S}_i$ with the line segment $\bar{S}_0\bar{S}_1$. When the condition $\left\lvert \bar{S}_0 - \bar{S}_1 \right\rvert \geq \sigma_0 + \sigma_1$ is satisfied, the solid classification circle $\bar{S}_i^{’}$ is internally tangent to the dotted circle $\bar{S}_i$ at $P_i$. Under this condition, the centers and radii of the classification circles, $\bar{S}_0^{’}$ and $\bar{S}_1^{’}$, can be written as:

\begin{subequations}
    \begin{equation}
        \bar{S}_0^{'} = \bar{S}_0 + {(\sigma_0^{'} - \sigma_0 )} \frac{\bar{S}_1 - \bar{S}_0}{\left\lvert \bar{S}_1 - \bar{S}_0 \right\rvert}, 
    \end{equation}
    \begin{equation}
        \bar{S}_1^{'} = \bar{S}_1 + {( \sigma_1^{'} - \sigma_1)} \frac{\bar{S}_0 - \bar{S}_1}{\left\lvert \bar{S}_1 - \bar{S}_0 \right\rvert}. 
    \end{equation}
\end{subequations}

Otherwise, circle $\bar{S}_i^{'}$ is externally tangent with circle $\bar{S}_j$ at $P_j$ ($i\neq j$),  and $\bar{S}_0^{'}$ and $\bar{S}_1^{'}$ write

\begin{subequations}
    \begin{equation}
        \bar{S}_0^{'} = \bar{S}_1 + {(\sigma_1 + \sigma_0^{'} )} \frac{\bar{S}_1 - \bar{S}_0}{\left\lvert \bar{S}_1 - \bar{S}_0 \right\rvert}, 
    \end{equation}
    \begin{equation}
        \bar{S}_1^{'} = \bar{S}_0 + {( \sigma_1^{'} + \sigma_0)} \frac{\bar{S}_0 - \bar{S}_1}{\left\lvert \bar{S}_1 - \bar{S}_0 \right\rvert}. 
    \end{equation}
\end{subequations}

During a bitstring measurement, if the IQ signal of a qubit falls within the circle $\bar{S}_0^{’}$ or $\bar{S}_1^{’}$, the qubit is classified as being in state $\ket{0}$ or $\ket{1}$, respectively. Otherwise, the measurement is discarded. As a result of discarding some experimental data, the total number of samples is reduced. To increase the number of usable experimental samples, we set $\sigma_i^{’}$ significantly larger than $\sigma_i$.

\subsubsection{Readout effects on qubits}

It is well known that an AC-Stark shift occurs in qubits when a readout microwave is applied. In this scenario, the frequency of a qubit may shift closer to the frequencies of other qubits or two-level systems (TLS), potentially introducing unwanted interactions during qubit readout. To mitigate these interactions and improve readout fidelity, the amplitude of the readout microwave can be adjusted to reduce the AC-Stark effect, or a Z bias can be applied to the qubits to tune their frequencies apart from each other during measurements.

\begin{figure}[H]
    \centering
    \begin{tikzpicture}
    \draw[elegant,orange,fill=white,domain=-\num+2.32-2.64:\num+3.02-2.64] plot(\x,{0.25*sin(50*\x r)-1});
    \draw[->] (2.5-2.64,-0.5)--(2.5-2.64,-0.75);
    \draw[->] (2.5-2.64,-1.5)--(2.5-2.64,-1.25);
    \node at (2.5-2.64,-0.5)[right]{$\Omega_\mathrm{R}$};
    
    \node at (0-2.64,-1.25)[left]{$Q_1$};
    \draw (0.5-2.64,-1)--(0.75-2.64,-1);\draw (0.75-2.64,-1)--(1.7-2.64,-1);
    \draw (3.65-2.64,-1)--(4.2-2.64,-1);\draw (4.8-2.64,-1)--(5.3-2.64,-1);
    \node at (0.2-2.64,-1) {$\ket{0}$}; 
    \node[draw] (B) at (4.5-2.64,-1) {$M$};

    \draw[blue] (0.5-2.64,-1.5)--(1.7-2.64,-1.5);
    \draw[blue] (3.65-2.64,-1.5)--(4.2-2.64,-1.5);\draw[blue] (4.8-2.64,-1.5)--(5.3-2.64,-1.5);
    \draw[blue] (1.7-2.64,-2.2)--(3.65-2.64,-2.2);\draw[blue] (4.2-2.64,-2.2)--(4.8-2.64,-2.2);
    \draw[blue] (1.7-2.64,-1.5)--(1.7-2.64,-2.2);\draw[blue] (3.65-2.64,-1.5)--(3.65-2.64,-2.2);\draw[blue] (4.2-2.64,-1.5)--(4.2-2.64,-2.2);
    \draw[blue] (4.8-2.64,-1.5)--(4.8-2.64,-2.2);
    \draw[<->] (2-2.64,-2.2)--node[below,sloped] {$Z_{1}$}(2-2.64,-1.5);
    \draw[dotted] (1.5-2.64,-1.5)-- (2.3-2.64,-1.5);

    \node at (0-2.64,-3.25)[left]{$Q_2$};
    \draw (0.5-2.64,-3)--(4.2-2.64,-3);\draw (4.8-2.64,-3)--(5.3-2.64,-3);
    \node at (0.2-2.64,-3) {$\ket{0}$}; 
    \node[draw] (B) at (4.5-2.64,-3) {$M$};
    \node[draw,fill=white] (B) at (1-2.64,-3) {$X$};
    \draw[blue] (0.5-2.64,-3.5)--(4.2-2.64,-3.5);\draw [blue](4.8-2.64,-3.5)--(5.3-2.64,-3.5);
    \draw[blue] (4.2-2.64,-3.5)--(4.2-2.64,-4.2); \draw[blue] (4.8-2.64,-3.5)--(4.8-2.64,-4.2);
    \draw[blue] (4.2-2.64,-4.2)--(4.8-2.64,-4.2);
    \draw[<->] (3.8-2.64,-4.2)--node[above,sloped] {$Z_{2 }$}(3.8-2.64,-3.5);
    \draw[dotted] (3.5-2.64,-4.2)-- (4.2-2.64,-4.2);
    \node (B) at (-1.6-2.7,-0.5) {\text{(a)}};
    
    \end{tikzpicture}
    \begin{tikzpicture}
        \node[anchor=south west,inner sep=0] (image) at (-1.5,0) {\includegraphics[width=0.45\textwidth]{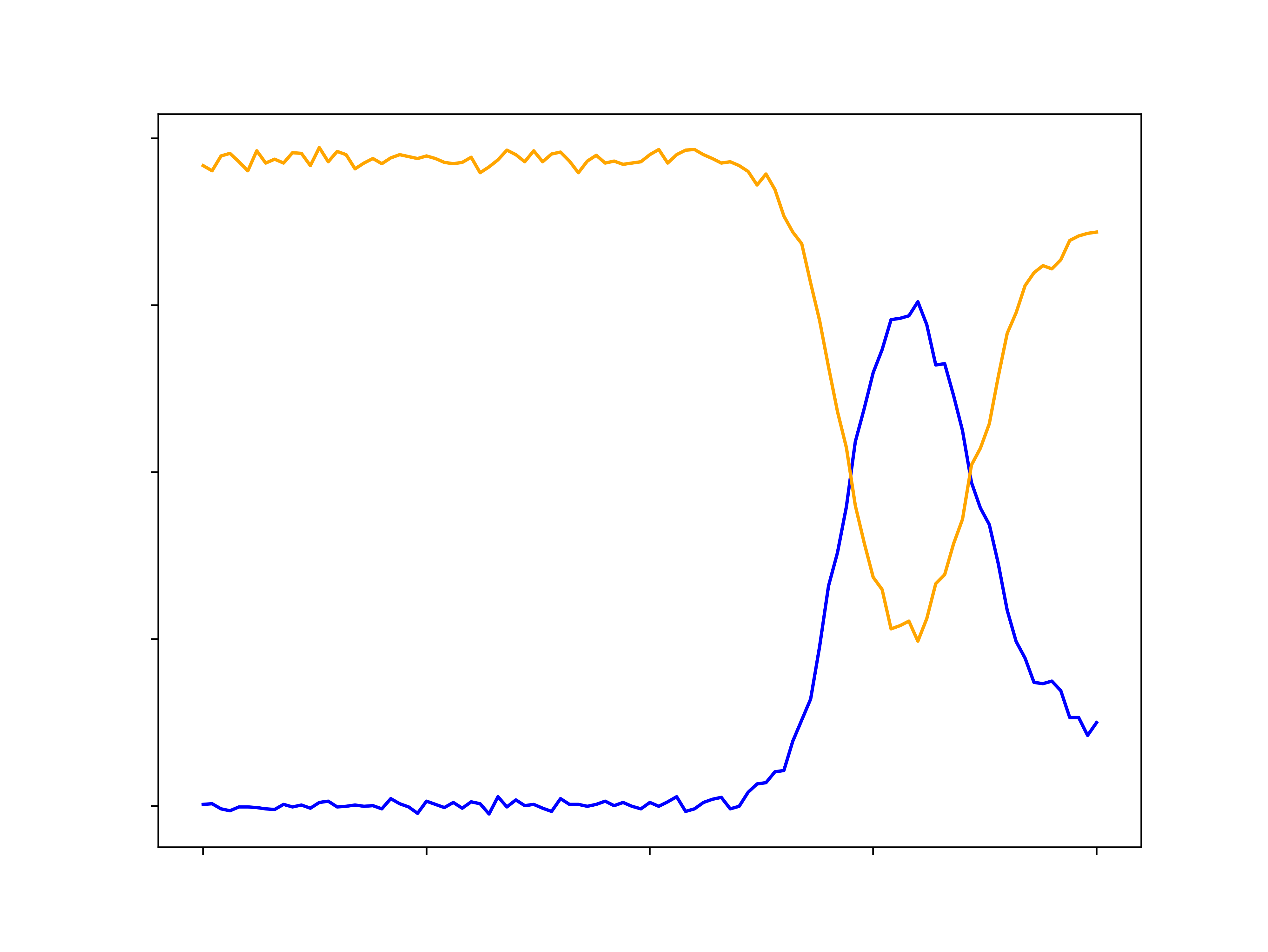}};
        \draw[blue] (5.3-5.6,4.5-0.8)--(5.8-5.6,4.5-0.8);
        \node[color=blue] (B) at (6-5.5,4.5-0.8) {$Q_1$};

        \draw[orange] (5.3-5.6,4-0.8)--(5.8-5.6,4-0.8);
        \node[orange] (B) at (6-5.5,4-0.8) {$Q_2$};

        \node (B) at (-0.2,0.4) {$0$};
        \node (B) at (1.2,0.4) {$0.03$};
        \node (B) at (2.6,0.4) {$0.06$};
        \node (B) at (4,0.4) {$0.09$};
        \node (B) at (5.4,0.4) {$0.12$};
        \node (B) at (2.6,-0.1) {$\Omega_\mathrm{R}$ (a.u.)};

        \node (B) at (-0.95,0.95) {$0$};
        \node (B) at (-0.95,2) {$0.25$};
        \node (B) at (-0.95,2+1.05) {$0.5$};
        \node (B) at (-0.95,2+2.1) {$0.75$};
        \node (B) at (-0.95,2+1.05*3) {$1$};

        \node[sloped] (B) at (-1.6,2+1.05) {\rotatebox{90}{$P_1$}};
        \node (B) at (-1.,5.7) {\text{(b)}};
    
    \end{tikzpicture}
    \caption{\textbf{Measurement of Readout effects on qubits}. 
    As shown in \textbf{a}, resonate drive (yellow pulse) with amplitude $\Omega_\mathrm{R}$ is applied on the readout resonator of $Q_1$ before measurement. During driving and measurement, Z bias $Z_{1 }$ and $Z_{2}$ are applied on $Q_1$ and $Q_2$, respectively. Then the $P_1$  of the two qubits are measured and shown in \textbf{b}. 
    }
    \label{Readeffect}
\end{figure}

As shown in Fig.~\ref{Readeffect}\textbf{a}, we utilize a pulse sequence to determine the amplitude ($R_\mathrm{amp}$) of the readout microwave and the Z biases ($Z_{1}$, $Z_{2}$) of the qubits. This pulse sequence can also be applied to specific qubits, such as $Q_2$ in Fig.~\ref{Readeffect}\textbf{a}, to identify the Z bias settings that yield optimal readout performance. 

Since multiple readouts of ancilla qubits are involved in our experimental circuit, it is necessary to examine the effects of these readouts on the data qubits. To this end, we resonantly drive the readout resonator of $Q_1$ with varying $R_\mathrm{amp}$ and measure the probabilities ($P_1$) of $Q_1$ and $Q_2$ remaining in their respective states. The results, shown in Fig.~\ref{Readeffect}\textbf{b}, indicate that tuning the readout amplitude of $Q_1$ influences the state exchange between $Q_1$ and $Q_2$. The appropriate $R_\mathrm{amp}$ for subsequent experiments should lie within a range where $P_1$ of $Q_2$ remains near zero, ensuring reliable qubit state readouts.

After completing these calibrations, we recheck the readout effects when only one qubit is excited prior to measurement. The IQ signal distributions of all qubits on the complex plane are then measured and analyzed to characterize these effects.

\subsubsection{Readout fidelity and decoherence time of qubits}

% \begin{figure}[H]
%     \centering
%     \begin{tikzpicture}
%     \node[fill=white] (B) at (-0.7,1){$\textbf{a}$};
%     \draw (0.2,0)--(7,0);
%     \node[fill=white] (B) at (0,0)[left]{$Q_a$};
%     \node[fill=white] (B) at (0.2,0) {$\ket{0}$}; 
    
%     \node[draw,fill=white] (B) at (1.2,0) {$I/X$};
%     \node[draw,fill=white] (B) at (2.5,0) {$M$};
%     \node[draw,fill=white] (B) at (3.8,0) {$M$};
%     \node[fill=white] (B) at (5.1,0) {$\cdots$};
%     \node[draw,fill=white] (B) at (6.4,0) {$M$};
%     \draw[<->] (3.8,0.4)--node[above,sloped] {$\tau=3.2\mu s$}(2.5,0.4);
%     \draw[dotted] (3.8,0.2)-- (3.8,0.6);
%     \draw[dotted] (2.5,0.2)-- (2.5,0.6);

%     \draw[<->] (2.5,-0.4)--node[below,sloped] {$N\times M$}(6.4,-0.4);
%     \draw[dotted] (6.4,-0.2)-- (6.4,-0.6);
%     \draw[dotted] (2.5,-0.2)-- (2.5,-0.6);
    
%     \end{tikzpicture}
%     \begin{tikzpicture}
        
%         \node[anchor=south west,inner sep=0] (image) at (-0.4,0.6) {\includegraphics[width=0.4\textwidth]{figures/sup/QND.png}};
%         \node[fill=white] (B) at (-0.7,4){$\textbf{b}$};
    
%     \end{tikzpicture}
%     \caption{\textbf{QND test of anccilla qubit}. As shown in \textbf{a}, we prepare anccilla qubit ($Q_a$) in an initial state ($\ket{0}$ or $\ket{1}$) and measure the probability of the initial state $N$ times every 3.2 $\mu s$. Then the results are show in \textbf{b}.  }
%     \label{QND}
    
% \end{figure}

\begin{figure}[H]
    \centering
    \begin{tikzpicture}
        \node[anchor=south west,inner sep=0] (image) at (-1,0) {\includegraphics[width=0.4\textwidth]{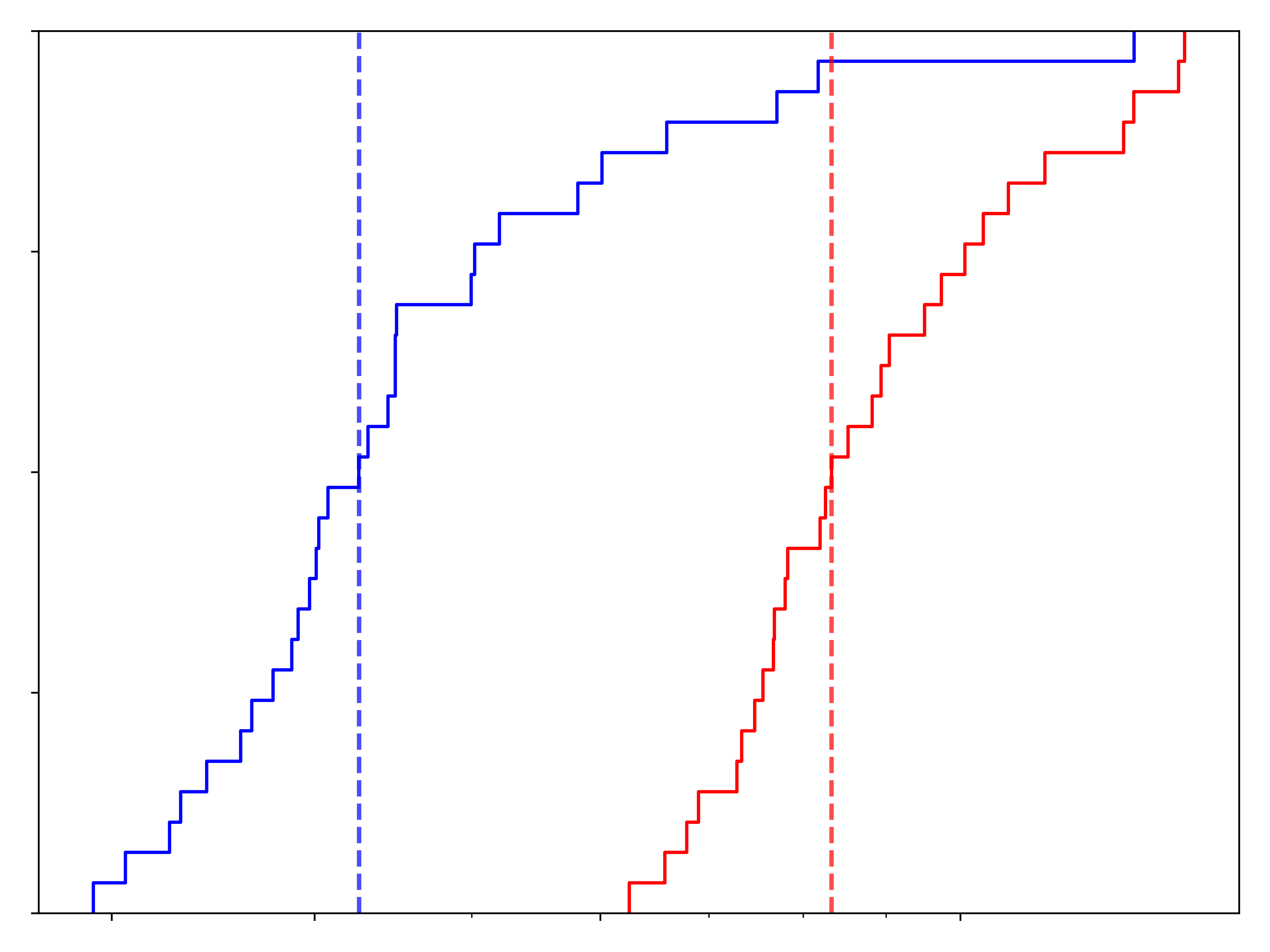}};
        \draw[blue] (-0.6,4.8)--(-0.0,4.8);
        \node[color=blue,draw opacity=0. ] (B) at (0.3,4.8) {$\ket{0}$};

        \draw[red] (-0.6,4.3)--(0,4.3);
        \node[red] (B) at (0.3,4.3) {$\ket{1}$};

        \node (B) at (0,-0.1) {$3\times10^{-2}$};
        % \node (B) at (1.,-0.1) {$4x10^{-2}$};
        \node (B) at (2.5,-0.1) {$6\times10^{-2}$};
        \node (B) at (4.5,-0.1) {$10^{-1}$};
        \node (B) at (2.6,-0.6) {SPAM error};

        \node (B) at (-0.95-0.4,-0.55+0.95) {$0$};
        \node (B) at (-0.95-0.4,-0.45+2.05) {$0.25$};
        \node (B) at (-0.95-0.4,-0.3+2+1.05) {$0.5$};
        \node (B) at (-0.95-0.4,-0.2+2+2.1) {$0.75$};
        \node (B) at (-0.95-0.4,-0.1+2+1.05*3) {$1$};

        \node[sloped] (B) at (-1.6-0.4,-0.5+2+1.05) {\rotatebox{90}{Integrated histogram (ECDF)}};
        % \node (B) at (-1.,5.7) {\textbf{b}};
    
    \end{tikzpicture}
    \caption{\textbf{Integrated histograms of state prepared and measurement(SPAM) error}. Integrated histograms for SPAM errors of qubits' states, $\ket{0}$ (blue) and $\ket{1}$ (red). Their median (dashed lines) are 0.043 and 0.083, respectively. 
    }
    \label{R_F}
\end{figure}

\begin{figure}[H]
    \centering
    \begin{tikzpicture}
        \node[anchor=south west,inner sep=0] (image) at (-1,0) {\includegraphics[width=0.4\textwidth]{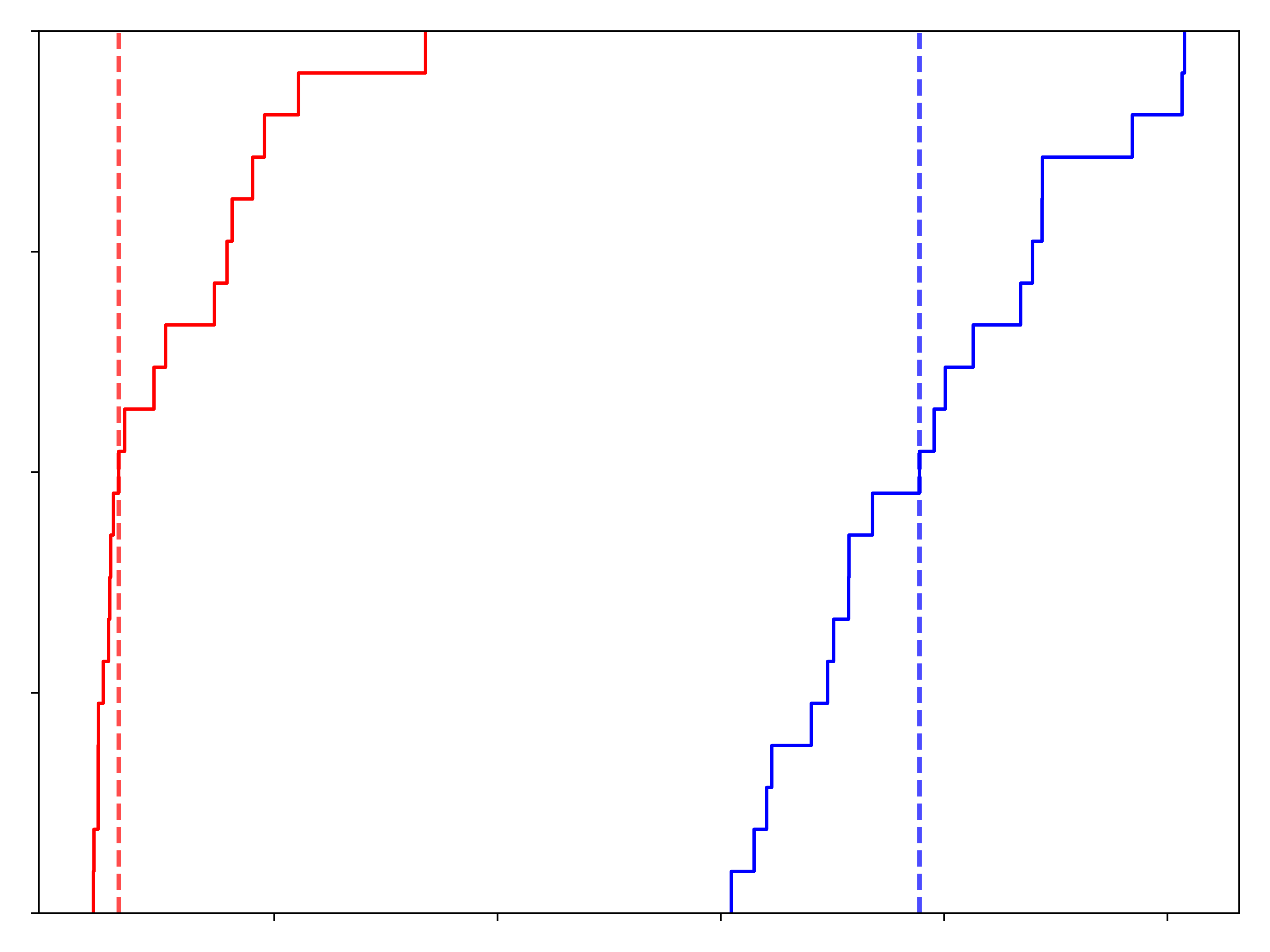}};
        \draw[blue] (0.9,3.5)--(1.5,3.5);
        \node[color=blue] (B) at (1.8,3.5) {$T_1$};
        \draw[red] (0.9,3)--(1.5,3);
        \node[color=red] (B) at (2.3,3) {$T_{2,\mathrm{Ramsey}}$};
        
        % \draw[red] (-0.6,4.3)--(0,4.3);
        % \node[red] (B) at (0.3,4.3) {$T_2$};

        \node (B) at (0.5,-0.1) {$20$};
        \node (B) at (1.8,-0.1) {$40$};
        \node (B) at (3.1,-0.1) {$60$};
        \node (B) at (4.4,-0.1) {$80$};
        \node (B) at (5.75,-0.1) {$100$};
        \node (B) at (2.6,-0.6) {Coherence time $(\mu\mathrm{s})$};

        \node (B) at (-0.95-0.3,-0.55+0.9) {$0$};
        \node (B) at (-0.95-0.3,-0.45+1.95) {$0.25$};
        \node (B) at (-0.95-0.3,-0.3+2+1.05) {$0.5$};
        \node (B) at (-0.95-0.3,-0.2+2+2.1) {$0.75$};
        \node (B) at (-0.95-0.3,-0.1+2+1.05*3) {$1$};

        \node[sloped] (B) at (-1.6-0.4,-0.5+2+1.05) {\rotatebox{90}{Integrated histogram (ECDF)}};
        % \node (B) at (-1.,5.7) {\textbf{b}};
    
    \end{tikzpicture}
    \caption{\textbf{Integrated histograms of coherence time}. Integrated histograms for coherence time of qubits, $T_1$ (blue) and $T_{2,\mathrm{Ramsey}}$(red). Their median (dashed lines) are 77.8 $\mu $s and 6.1 $\mu $s. 
    }
    \label{T1T2}
\end{figure}

Due to thermalization and the long resonator ringdown times ($1/\kappa$), qubits cannot be effectively initialized without resetting their states, and the readout duration cannot be shortened, resulting in a readout time of 2 $\mu$s. The readout errors (post-selected using the method described in subsection A) are shown in Fig.~\ref{R_F}, while the coherence times are presented in Fig.~\ref{T1T2}.

% the quantum non demolition measurement pulse sequence and results In Fig.~\ref{QND}, 

\subsection{Calibration of quantum gates}

In our experiment, single-qubit gates are implemented with the qubits and couplers maintained at their idle points, minimizing the impact of XY and ZZ couplings on the gate preparation. The controlled-Z (CZ) gate between two qubits (e.g., $Q_1$ and $Q_2$) is realized by tuning the frequency of the coupler via Z bias, while the qubit frequencies remain fixed at their idle points~\cite{Huikai}.

The parameters of the CZ gate depend on the Z bias applied to the coupler, which follows an inverted hyperbolic cosine function. During the execution of the CZ gate, state leakage out of the computational subspace must be suppressed, and the conditional phase should accumulate to $\pi$. Additionally, the distortion and delay in the coupler’s Z bias are calibrated using the method described in Ref.~\cite{zhao_quantum_2024}.

\subsubsection{Idle point of coupler}

\begin{figure}[H]
    \centering
    \begin{tikzpicture}
        \node at (-0.5,0.5)[left]{\text{(a)}};
        \node at (-0.1,0)[left]{$C$};
        \draw (0.5,0)--(1.3,0);\draw (1.3,0)--(1.3,-0.5);\draw (1.3,-0.5)--(2.6,-0.5);
        \draw (2.6,-0.5)--(2.6,0);\draw (2.6,0)--(3.3,0);\draw (3.3,0)--(3.3,-0.5);
        \draw (3.3,-0.5)--(4.6,-0.5);\draw (4.6,-0.5)--(4.6,0); \draw (4.6,0)--(7,0);
        \node at (0.2,0) {$\ket{0}$}; 
        \draw[<->] (1.3,-0.25)--node[above,sloped] {$\tau$}(2.6,-0.25);
        \draw[<->] (3.3,-0.25)--node[above,sloped] {$\tau$}(4.6,-0.25);
        \draw[<->] (5,-0.5)--node[below,sloped] {$Z_C$}(5,0);
        \draw[densely dotted] (4.6,-0.5)--(5.2,-0.5);
        \node (B) at (2.,0.3) {$U_{\phi}$};

        \node at (0,-1)[left]{$Q_1$};
        \draw (0.5,-1)--(0.75,-1);\draw (1.25,-1)--(2.75,-1);\draw (3.25,-1)--(4.75,-1);
        \draw (5.27,-1)--(6.2,-1);\draw (6.8,-1)--(7,-1);
        \node at (0.2,-1) {$\ket{0}$}; 
        \node[draw] (B) at (1,-1) {$\frac{X}{2}$};
        \node[draw] (B) at (3,-1) {$X$};
        \node[draw] (B) at (5,-1) {$\frac{X}{2}$};
        \node[draw] (B) at (6.5,-1) {$M$};

        \node at (0,-2)[left]{$Q_2$};
        \draw (0.5,-2)--(2.75,-2);\draw (3.27,-2)--(7,-2);
        \node at (0.2,-2) {$\ket{0}$}; 
        \node[draw] (B) at (3,-2) {$X$};
    \end{tikzpicture}
    \begin{tikzpicture}
        \node[anchor=south west,inner sep=0] (image) at (-0.7,0.6) {\includegraphics[width=0.38\textwidth]{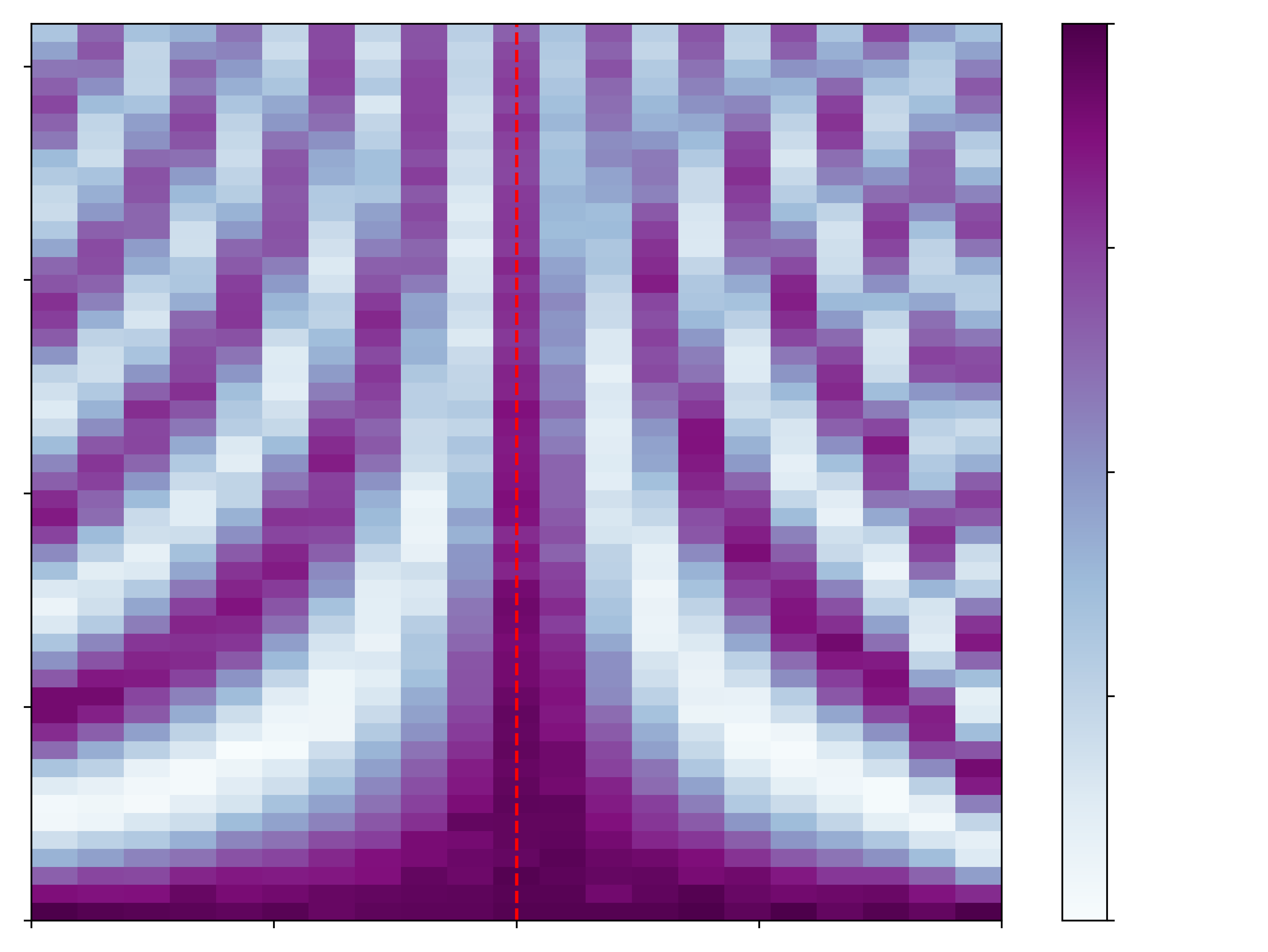}};

        \node (B) at (-0.5,0.4) {$-0.1$};
        \node (B) at (0.75,0.4) {$-0.05$};
        \node (B) at (2.1,0.4) {$0$};
        \node (B) at (3.4,0.4) {$0.05$};
        \node (B) at (4.7,0.4) {$0.1$};
        \node (B) at (2.1,-0.1) {$Z_{C}$ (a.u.)};

        \node (B) at (-0.95,0.8) {$0$};
        \node (B) at (-0.95,1.95) {$3.5$};
        \node (B) at (-0.95,3.05) {$7$};
        \node (B) at (-0.95,4.25) {$10.5$};
        \node (B) at (-0.95,5.35) {$14$};
        \node[sloped] (B) at (-1.4,2+1.05) {\rotatebox{90}{$\tau(\mu\mathrm{s})$}};
        \node (B) at (-1.4,6) {\text{(b)}};

        \node (B) at (5.7,0.8) {$0$};
        \node (B) at (5.7,1.95) {$0.25$};
        \node (B) at (5.7,3.15) {$0.5$};
        \node (B) at (5.7,4.4) {$0.75$};
        \node (B) at (5.7,5.56) {$1$};

        \node[sloped] (B) at (6.3,3.15) {\rotatebox{90}{$P_1$}};
    
    \end{tikzpicture}
    \caption{\textbf{Measurement of conditional phase and ZZ coupling}.
    \textbf{a} shows an echo type sequence to measure conditional phase. Among which, $\tau$ and $Z_C$ are length and strength of Z bias of coupler ($C$), respectively. 
    \textbf{b} shows the probability ($P_1$) of $Q_1$ in state $\ket{1}$ varying with $\tau$ and $Z_C$. 
    } 
    \label{ZZ_vs_C}
\end{figure}

We position the two qubits ($Q_1$, $Q_2$) at their idle points, where their large detuning minimizes the influence of XY coupling on single-qubit gates. Additionally, the coupler ($C$) is set to its idle point to minimize ZZ coupling between the qubits. As shown in Fig.~\ref{ZZ_vs_C}\textbf{a}, an echo-type pulse sequence is used to measure the conditional phase of $Q_1$ and $Q_2$. Tuning the coupler’s frequency can effectively be modeled as applying a CPHASE gate ($U_{\phi}$) on the two qubits, provided that state leakage and qubit exchange are negligible. The operation $U_{\phi}$ can be expressed as:

\begin{subequations}
    \begin{equation}
    U_{\phi}= \mathrm{diag}(1, e^{-i\phi_2}, e^{-i\phi_1}, e^{-i\phi_{12}}),
\end{equation}
\begin{equation}
    \phi_{12} = \phi_{1} + \phi_{2} + \varphi,
\end{equation}
\end{subequations}

where $\phi_{1}$ and $ \phi_{2}$ are the local Z rotation phases of $Q_1$ and $Q_2$, respectively, and $ \varphi$ is conditional phase. Considering the effect of tuning the coupler's frequency on the two qubits as $U_{\phi}$, the exciting probability $P_1$ of $Q_1$ can be derived as:

\begin{equation}
    P_1 = \frac{1}{2}(1 - \cos\varphi),   
\end{equation}

The ZZ coupling is given by $\alpha_{zz} = \varphi / \tau$, which corresponds to the oscillation frequency of $P_1$. By varying $\tau$ and the coupler bias $Z_c$, we measure the probability $P_1$ of $Q_1$, as shown in Fig.~\ref{ZZ_vs_C}\textbf{b}. The idle bias of the coupler is determined where $\alpha_{zz}$ reaches its minimum value, indicated by the red dashed line in Fig.~\ref{ZZ_vs_C}\textbf{b}.

\subsubsection{Calibration of CZ gate}

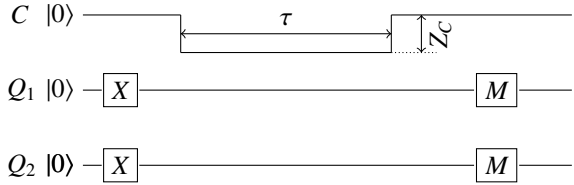
\begin{figure}[H]
    \centering
    \begin{tikzpicture}
        \node at (-0.1,0)[left]{$C$};
        \draw (0.5,0)--(1.8,0);\draw (1.8,0)--(1.8,-0.5);\draw (1.8,-0.5)--(4.6,-0.5);
        \draw (4.6,-0.5)--(4.6,0); \draw (4.6,0)--(7,0);
        \node at (0.2,0) {$\ket{0}$}; 
        \draw[<->] (1.8,-0.25)--node[above,sloped] {$\tau$}(4.6,-0.25);
        \draw[<->] (5,-0.5)--node[below,sloped] {$Z_{C}$}(5,0);
        \draw[densely dotted] (4.6,-0.5)--(5.2,-0.5);

        \node at (0,-1)[left]{$Q_1$};
        \draw (0.5,-1)--(0.75,-1);\draw (1.25,-1)--(6.2-0.5,-1);\draw (6.8-0.5,-1)--(7,-1);
        \node at (0.2,-1) {$\ket{0}$}; 
        \node[draw] (B) at (1,-1) {$X$};
        \node[draw] (B) at (6,-1) {$M$};

        \node at (0,-2)[left]{$Q_2$};
        \draw (0.5,-2)--(0.75,-2);\draw (1.25,-2)--(6.2-0.5,-2);\draw (6.8-0.5,-2)--(7,-2);
        \node at (0.2,-2) {$\ket{0}$}; 
        \node[draw] (B) at (1,-2) {$X$};
        \node[draw] (B) at (6,-2) {$M$};
        \node at (0.2,-2) {$\ket{0}$}; 

    \end{tikzpicture}
    \caption{\textbf{Characterization of state leakage}. 
    Using the pulse sequence can measure the joint probability $P_{11}$ to characterize state leakage. 
    $\tau$ and $Z_C$ are length and amplitude of Z bias of coupler ($C$), respectively. } 
    \label{LKG_P}
\end{figure}

\begin{figure}[H]
    \centering
    \begin{tikzpicture}
        \node[anchor=south west,inner sep=0] (image) at (-0.7,0.6) {\includegraphics[width=0.38\textwidth]{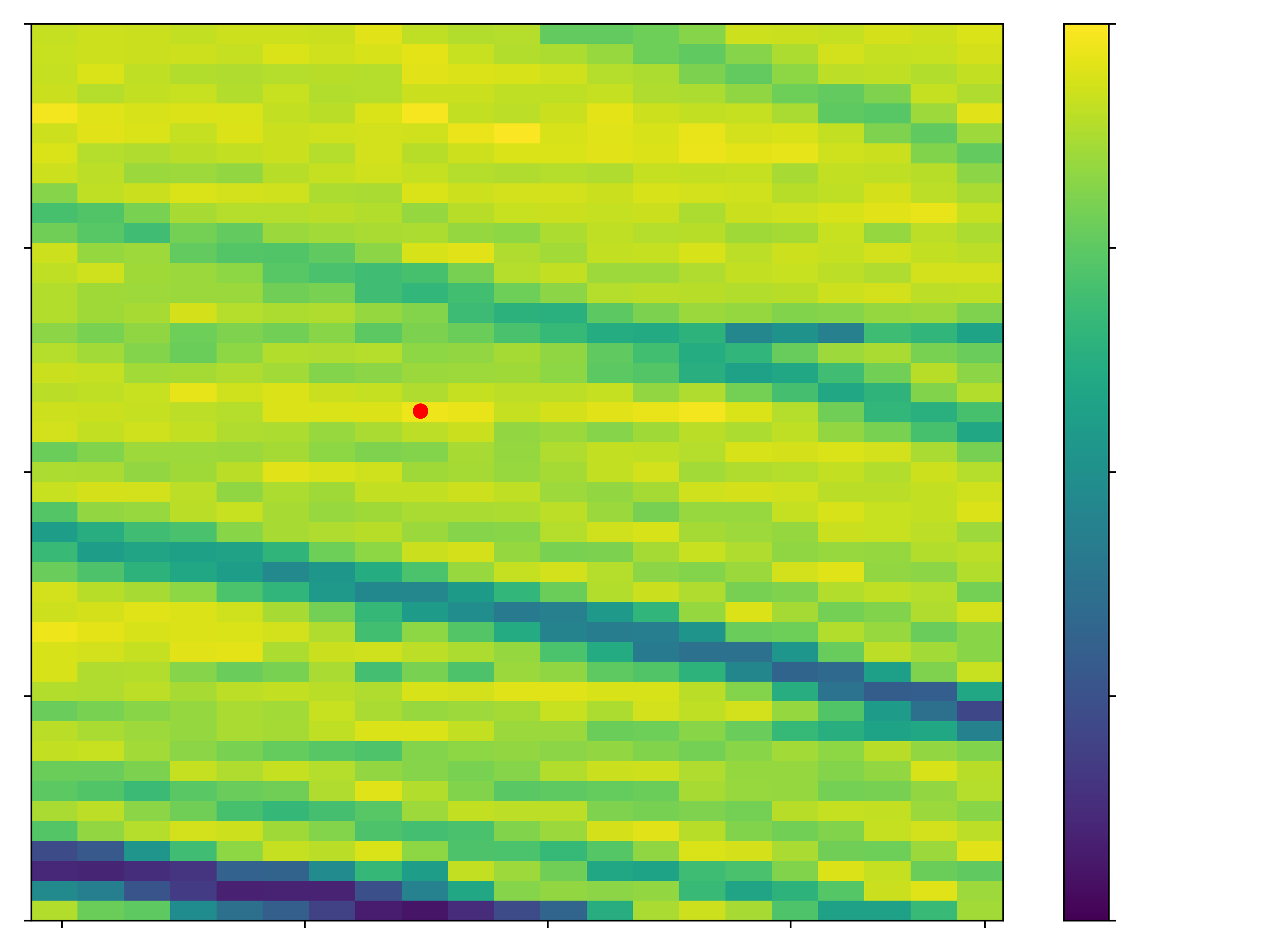}};

        \node (B) at (-0.5,0.4) {$0.135$};
        \node (B) at (0.85,0.4) {$0.140$};
        \node (B) at (2.2,0.4) {$0.145$};
        \node (B) at (3.5,0.4) {$0.150$};
        \node (B) at (4.6,0.4) {$0.154$};
        \node (B) at (2.1,-0.1) {$Z_C$ (a.u.)};

        \node (B) at (-0.95,0.8) {$80$};
        \node (B) at (-0.95,1.95) {$102$};
        \node (B) at (-0.95,3.15) {$124$};
        \node (B) at (-0.95,4.35) {$146$};
        \node (B) at (-0.95,5.55) {$168$};
        \node[sloped] (B) at (-1.4,2+1.05) {\rotatebox{90}{$\tau (\mathrm{ns})$}};
        \node (B) at (-1.4,6) {\text{(a)}};

        \node (B) at (5.7,0.8) {$0$};
        \node (B) at (5.7,1.95) {$0.25$};
        \node (B) at (5.7,3.15) {$0.5$};
        \node (B) at (5.7,4.4) {$0.75$};
        \node (B) at (5.7,5.56) {$1$};

        \node[sloped] (B) at (6.3,3.15) {\rotatebox{90}{$P_{11}$}};
    
    \end{tikzpicture}
    \begin{tikzpicture}
        \node[anchor=south west,inner sep=0] (image) at (-0.7,0.6) {\includegraphics[width=0.38\textwidth]{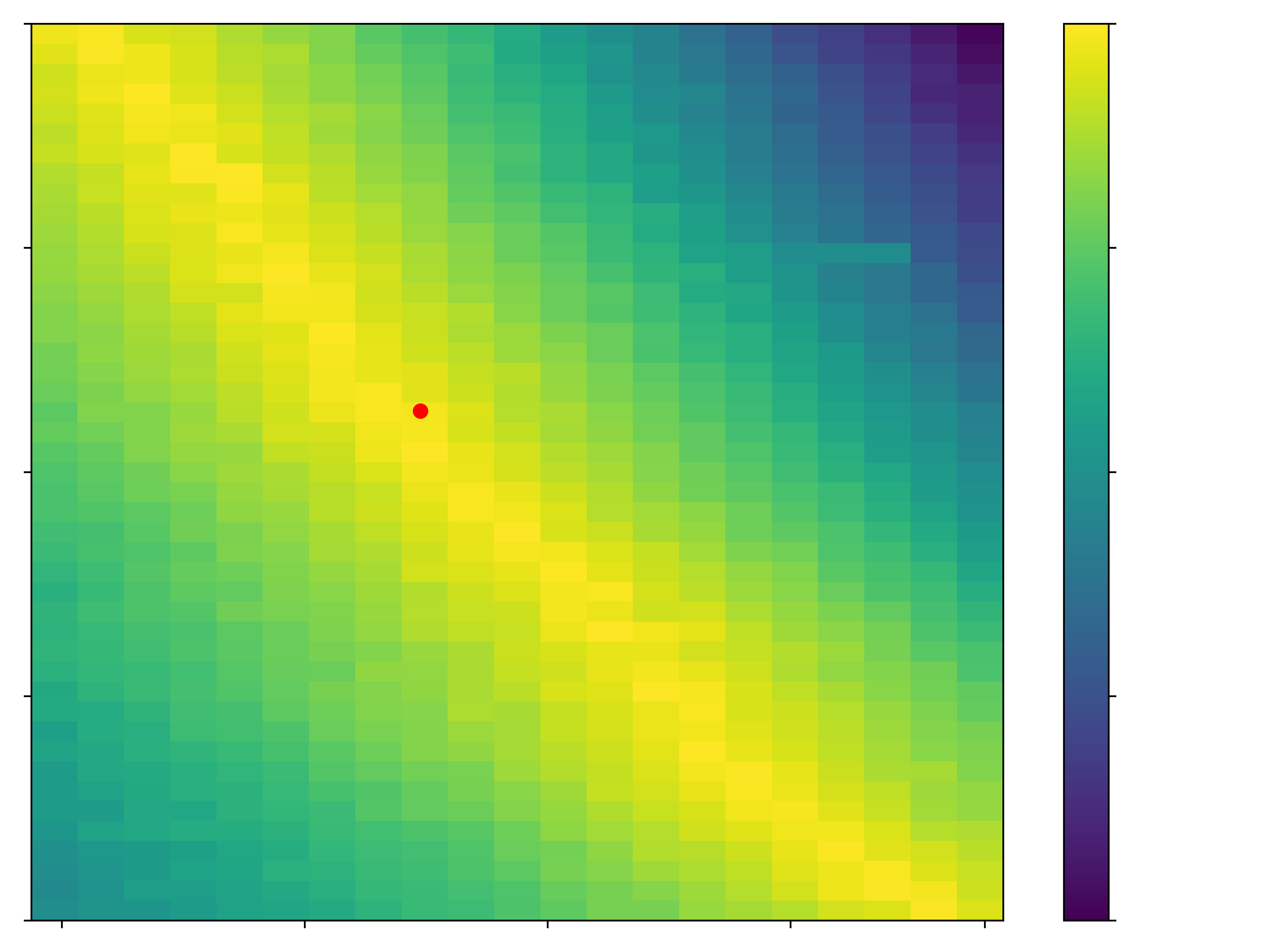}};

        \node (B) at (-0.5,0.4) {$0.135$};
        \node (B) at (0.85,0.4) {$0.140$};
        \node (B) at (2.2,0.4) {$0.145$};
        \node (B) at (3.5,0.4) {$0.150$};
        \node (B) at (4.6,0.4) {$0.154$};
        \node (B) at (2.1,-0.1) {$Z_C$ (a.u.)};

        \node (B) at (-0.95,0.8) {$80$};
        \node (B) at (-0.95,1.95) {$102$};
        \node (B) at (-0.95,3.15) {$124$};
        \node (B) at (-0.95,4.35) {$146$};
        \node (B) at (-0.95,5.55) {$168$};
        \node[sloped] (B) at (-1.4,2+1.05) {\rotatebox{90}{$\tau(\mu\mathrm{s})$}};
        \node (B) at (-1.4,6) {\text{(b)}};

        \node (B) at (5.7,0.8) {$0$};
        \node (B) at (5.7,1.95) {${\pi}/{4}$};
        \node (B) at (5.7,3.15) {${\pi}/{2}$};
        \node (B) at (5.7,4.4) {${3\pi}/{4}$};
        \node (B) at (5.7,5.56) {$\pi$};

        \node[sloped] (B) at (6.3,3.15) {\rotatebox{90}{$\varphi$}};
    
    \end{tikzpicture}
    \caption{\textbf{State leakage and conditional phase}. 
    \textbf{a} and \textbf{b} show the experiment results of $P_{11}$ and $\varphi$ varying with $\tau$ and $Z_C$ shown in Fig.~\ref{ZZ_vs_C}\textbf{a} and  Fig.~\ref{LKG_P}, respectively. } 
    \label{leakage}
\end{figure}

The parameters of the CZ gate include the Z bias of the coupler, the pulse length ($\tau$), and the amplitude ($Z_C$). To realize a CZ gate, these parameters must suppress state leakage to ensure that a $U_{\phi}$ gate is applied to the two qubits while simultaneously accumulating a conditional phase of $\pi$. We implement the pulse sequence shown in Fig.~\ref{LKG_P} on the two qubits and the coupler, measuring the joint probability ($P_{11}$) of the two qubits to experimentally characterize state leakage. Once leakage occurs, the value of $P_{11}$ will decrease. Fig.~\ref{leakage}\textbf{a} shows the variation of $P_{11}$ with the experimental parameters $Z_C$ and $\tau$ of the CZ gate. The optimal parameters must satisfy the condition $P_{11} \approx 1$ to successfully achieve a $U_{\phi}$ gate.

The circuit used for conditional phase calibration is similar to Fig.~\ref{ZZ_vs_C}\textbf{a}, except this time we first bias the qubits and the coupler at their idle biases, and then apply a Z pulse with an amplitude of $Z_C$ to the coupler at the position where the CZ gate is implemented. To obtain a more precise measurement of $\varphi$, we replace the second $\frac{X}{2}$ pulse on $Q_1$ with a tomography sequence and perform a tomography measurement. Conditional phase $\varphi$ extract from tomography are shown in Fig.~\ref{leakage}\textbf{b}. The parameters $\tau$ and $Z_C$ for the CZ gate must simultaneously satisfy $P_{11} = 1$ and $\varphi = \pi$. These optimal values, labeled as $\tau_\mathrm{cz}$ and $Z_\mathrm{cz}$, respectively, are marked as red points in Fig.~\ref{leakage}.

\begin{figure}[H]
    \centering
    \begin{tikzpicture}
        \node at (-0.1,0)[left]{$C$};
        \node at (0.2,0) {$\ket{0}$}; 
        \node at (-0.9,0.5)[left]{\text{(a)}};
        \draw (0.5,0)--(1.3,0);\draw (1.3,0)--(1.3,-0.5);\draw (1.3,-0.5)--(2,-0.5);
        \draw (2,-0.5)--(2,0); \draw (2,0)--(3.4,0);
        \draw[<->] (1.3,-0.25)--node[above,sloped] {$\tau_\mathrm{cz}$}(2,-0.25);
        \draw[<->] (2.1,-0.5)--node[below,sloped] {$Z_\mathrm{cz}$}(2.1,0);

        \draw (3.4,0)--(3.4,-0.5);\draw (3.4,-0.5)--(4.1,-0.5);
        \draw (4.1,-0.5)--(4.1,0); \draw (4.1,0)--(7,0);
        \draw[<->] (3.4,-0.25)--node[above,sloped] {$\tau_\mathrm{cz}$}(4.1,-0.25);
        \draw[<->] (4.2,-0.5)--node[below,sloped] {$Z_\mathrm{cz}$}(4.2,0);
        
        % \draw[densely dotted] (2,-0.5)--(3.1,-0.5);
        \node at (2,0.3)[above]{$\times 2N$};
        \node at (4.1,0.3)[above]{$\times 2N$};

        \node at (0,-1)[left]{$Q_1$};
        \draw (0.5,-1)--(0.65,-1);\draw (1.15,-1)--(2.06,-1);\draw (2.54,-1)--(2.73,-1);\draw (3.24,-1)--(3.93+0.26,-1);\draw (4.44+0.26,-1)--(5.05,-1);\draw (5.55,-1)--(5.9,-1);\draw (7-0.5,-1)--(7,-1);
        \node at (0.2,-1) {$\ket{0}$}; 
        \node[draw] (B) at (0.9,-1) {$\frac{X}{2}$};
        \node[draw] (B) at (2.3,-1) {$X$};
        \node[draw] (B) at (4.2+0.26,-1) {$X$};
        \node[draw] (B) at (3,-1) {$X$};
        \node[draw] (B) at (6.2,-1) {$M$};
        \node[draw] (B) at (5.3,-1) {$\frac{X}{2}$};

        \node at (0,-2)[left]{$Q_2$};
        \draw (0.5,-2)--(0.65,-2);\draw (1.15,-2)--(2.06,-2);\draw (2.54,-2)--(2.73,-2);\draw (3.24,-2)--(3.93+0.26,-2);\draw (4.44+0.26,-2)--(7,-2);
        \node at (0.2,-2) {$\ket{0}$}; 
        \node[draw] (B) at (2.3,-2) {$X$};
        \node[draw] (B) at (4.2+0.26,-2) {$X$};
        \node[draw] (B) at (3,-2) {$X$};
        \node at (0.2,-2) {$\ket{0}$}; 

        \draw[dashed] (1.25,0.3)--(1.25,-2.4);\draw[dashed] (2.65,0.3)--(2.65,-2.4);
        \draw[dashed] (1.25,0.3)--(2.65,0.3);\draw[dashed] (1.25,-2.4)--(2.65,-2.4);

        \draw[dashed] (3.33,0.3)--(3.33,-2.4);\draw[dashed] (4.83,0.3)--(4.83,-2.4);
        \draw[dashed] (3.33,0.3)--(4.83,0.3);\draw[dashed] (3.33,-2.4)--(4.83,-2.4);

    \end{tikzpicture}
    
    \begin{tikzpicture}
        \node[anchor=south west,inner sep=0] (image) at (-1.3,0.5) {\includegraphics[width=0.4\textwidth]{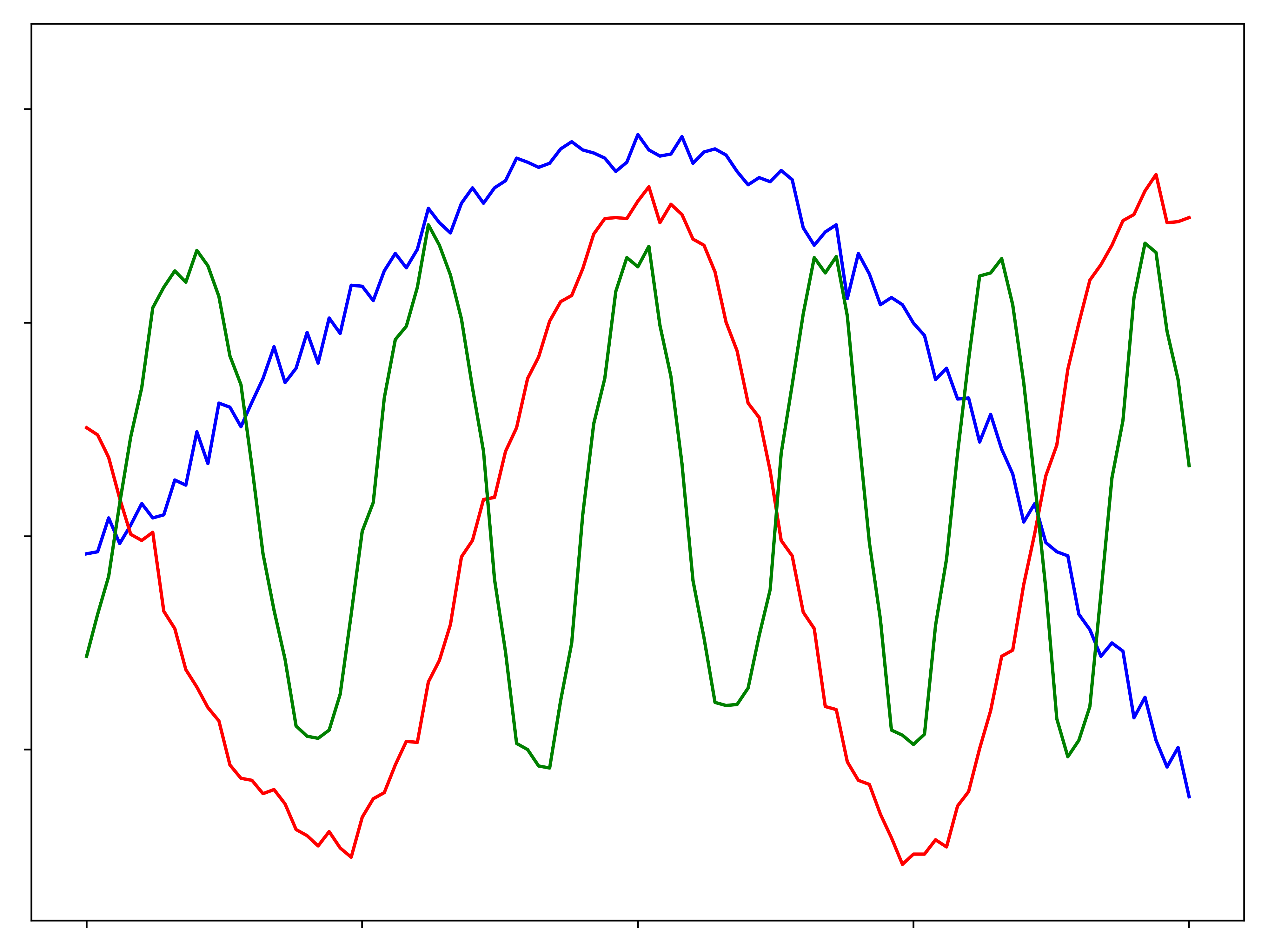}};
        \draw[blue] (5.3-5.6-0.7,4.5+1)--(5.8-5.6-0.8,4.5+1);
        \node[color=blue] (B) at (6-5.5-0.7,4.5+1) {N=1};
        
        \draw[green] (5.3-5.6+1.9,4.5+1)--(5.8-5.6+1.8,4.5+1);
        \node[green] (B) at (6-5.5+1.9,4.5+1) {N=9};

        \draw[red] (5.3-5.5 +0.5,4.5+1)--(5.8-5.6 +0.5,4.5+1);
        \node[color=red] (B) at (6-5.4 +0.5,4.5+1) {N=3};

        \node (B) at (-0.92,0.4) {$-0.005$};
        \node (B) at (0.7,0.4) {$-0.0025$};
        \node (B) at (2.325,0.4) {$0$};
        \node (B) at (3.9,0.4) {$0.0025$};
        \node (B) at (5.4,0.4) {$0.005$};
        \node (B) at (2.6,-0.1) {$Z_\mathrm{cz}$ (a.u.)};

        % \node (B) at (-0.95-0.7,0.95) {$0$};
        \node (B) at (-0.95-0.65,2-0.35) {$0.25$};
        \node (B) at (-0.95-0.65,2+0.9) {$0.5$};
        \node (B) at (-0.95-0.65,2+2.1) {$0.75$};
        \node (B) at (-0.95-0.65,2+3.3) {$1$};

        \node[sloped] (B) at (-1.6-0.65,2+1.05) {\rotatebox{90}{$P_1$}};
        \node (B) at (-1.6-0.65,6.5) {\text{(b)}};

        \draw[dashed] (6-5.5+1.8,1)--(6-5.5+1.8,5.4);
    
    \end{tikzpicture}
    \caption{\textbf{Optimization of coupler's Z bias of CZ gate}. 
    \textbf{a} shows the pulse train to optimize coupler's Z bias of CZ gate. We repeat the pulse sequence in the dashed box $2N$ times, measure the probability ($P_1$) of $Q_1$ and obtain the results shown in \textbf{b}.  } 
    \label{PulseTrain}
\end{figure}

Subsequently, we use a pulse train, as shown in Fig.~\ref{PulseTrain}\textbf{a}, to further optimize $Z_\mathrm{cz}$ by fine-tuning it and measuring the probability $P_1$ of $Q_1$, with the results shown in Fig.~\ref{PulseTrain}\textbf{b}. Considering the effect of the coupler’s Z bias on the two qubits as $U_{\phi}$, the probability $P_1$ of $Q_1$ can be expressed as:

\begin{equation}
    P_1 = \frac{1}{2}(1 + \cos{2N\varphi}). 
\end{equation}
As shown in  Fig.~\ref{PulseTrain}\textbf{b}, for different N, $P_1$ has a maximum if $\varphi=\pi$. 

\begin{figure}[H]
    \centering
    \begin{tikzpicture}
        \node at (-0.1,0)[left]{$C$};
        \node at (-0.9,0.5)[left]{\text{(a)}};
        \draw (0.5,0)--(1.8,0);\draw (1.8,0)--(1.8,-0.5);\draw (1.8,-0.5)--(3.5,-0.5);
        \draw (3.5,-0.5)--(3.5,0); \draw (3.5,0)--(7,0);
        \node at (0.2,0) {$\ket{0}$}; 
        \draw[<->] (1.8,-0.25)--node[above,sloped] {$\tau_\mathrm{cz}$}(3.5,-0.25);
        \draw[<->] (3.8,-0.5)--node[below,sloped] {$Z_\mathrm{cz}$}(3.8,0);
        \draw[densely dotted] (3.5,-0.5)--(4.1,-0.5);
        \node at (3.3,0.3)[above]{$\times N$};

        \node at (0,-1)[left]{$Q_i$};
        \draw (0.5,-1)--(0.75,-1);\draw (1.25,-1)--(4.55-0.5,-1);\draw (4.55,-1)--(5.05,-1);\draw (5.55,-1)--(5.9,-1);\draw (7-0.5,-1)--(7,-1);
        \node at (0.2,-1) {$\ket{0}$}; 
        \node[draw] (B) at (1,-1) {$\frac{X}{2}$};
        \node[draw,fill=white] (B) at (4.2,-1) {$Z_{\phi_i}$};
        \node[draw] (B) at (6.2,-1) {$M$};
        \node[draw] (B) at (5.3,-1) {$\frac{X}{2}$};

        \draw[dashed] (1.5,0.3)--(1.5,-1.6);\draw[dashed] (4.7,0.3)--(4.7,-1.6);
        \draw[dashed] (1.5,0.3)--(4.7,0.3);\draw[dashed] (1.5,-1.6)--(4.7,-1.6);

    \end{tikzpicture}
    \begin{tikzpicture}
        \node[anchor=south west,inner sep=0] (image) at (-1.3,0.5) {\includegraphics[width=0.4\textwidth]{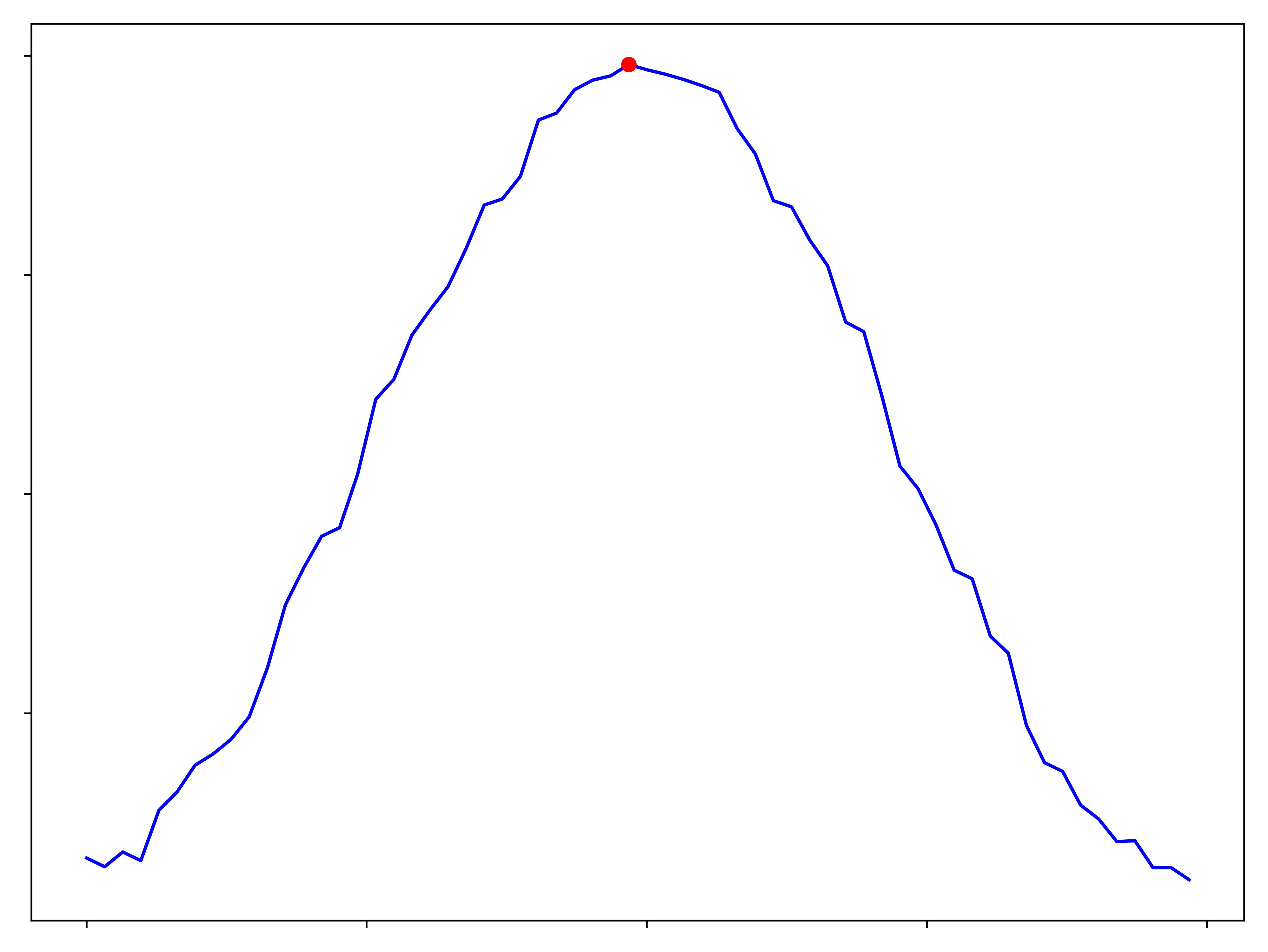}};

        \node (B) at (-0.92,0.4) {$-\pi$};
        \node (B) at (0.7,0.4) {$-\pi/2$};
        \node (B) at (2.38,0.4) {$0$};
        \node (B) at (4,0.4) {$\pi/2$};
        \node (B) at (5.55,0.4) {$\pi$};
        \node (B) at (2.6,-0.1) {$\phi_i$ (rad)};

        \node (B) at (-0.95-0.65,2-0.15) {$0.25$};
        \node (B) at (-0.95-0.65,2+1.1) {$0.5$};
        \node (B) at (-0.95-0.65,2+2.3) {$0.75$};
        \node (B) at (-0.95-0.65,2+3.6) {$1$};

        \node[sloped] (B) at (-1.6-0.65,2+1.05) {\rotatebox{90}{$P_1$}};
        \node (B) at (-1.6-0.65,6) {\text{(b)}};
    
    \end{tikzpicture}
    \caption{\textbf{Elimination of single-qubit phase}. 
    \textbf{a} shows a pulse sequence to eliminate single-qubit phase. 
    The pulse sequence in dashed box is repeated $N$ times and followed a probability ($P_1$) measurement of $Q_1$, and the measurement results are shown in \textbf{(b)}.
    } 
    \label{Dphase}
\end{figure}

To achieve a standard CZ gate, it is also necessary to measure and eliminate the local Z rotation phases $\phi_1$ and $\phi_2$. For this purpose, a virtual Z gate is employed to cancel out the local Z rotation phases. As shown in Fig.~\ref{Dphase}\textbf{a}, we apply a virtual Z gate ($Z_{\phi}$) on $Q_i$ during the execution of the CZ gate. The probability $P_1$ of $Q_i$ then takes the form

\begin{equation}
    P_1 = \frac{1}{2}(1 + \cos{N (\phi_i + \phi)}). 
\end{equation}

Here, $P_1$ reaches its maximum value when $Z_{\phi}$ cancels the single-qubit phase $\phi_i$, i.e., $\phi_i + \phi = 0$, as indicated by the red point in Fig.~\ref{Dphase}\textbf{b}.

\subsubsection{Leakage and fidelity of quantum gate}

\begin{figure}[H]
    \centering
    \begin{tikzpicture}
        \node[anchor=south west,inner sep=0] (image) at (-2,-0.5) {\includegraphics[width=0.5\textwidth]{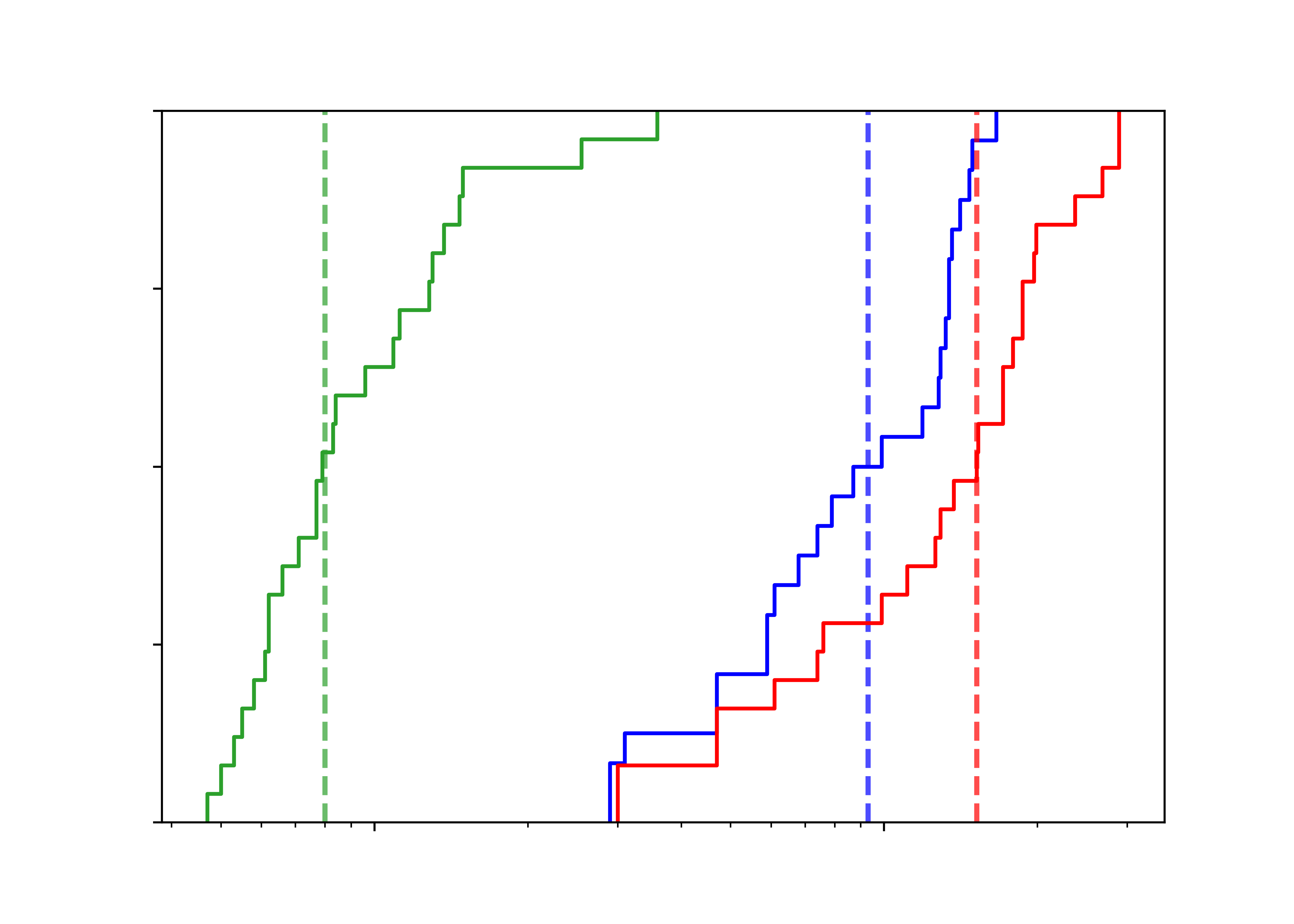}};
        \draw[green] (1.3,3.9)--(1.9,3.9);
        \node[color=green] (B) at (2.8,3.95) {parallel SG};

        \draw[red] (1.3,3.5)--(1.9,3.5);
        \node[color=red] (B) at (2.8,3.55) {parallel CZ};

        \draw[blue] (1.3,3.1)--(1.9,3.1);
        \node[color=blue] (B) at (2.8,3.15) {isolated CZ};

        % \node (B) at (-0.05,-0.1) {$0.0$};
        \node (B) at (0.8,-0.1) {$10^{-3}$};
        \node (B) at (2.5,-0.1) {$3\times10^{-3}$};
        \node (B) at (4.3,-0.1) {$10^{-2}$};
        \node (B) at (2.6,-0.6) {RB error rate};

        \node (B) at (-0.95-0.4,-0.65+0.95) {$0$};
        \node (B) at (-0.95-0.4,-0.55+2.) {$0.25$};
        \node (B) at (-0.95-0.4,-0.35+2+1.) {$0.5$};
        \node (B) at (-0.95-0.4,-0.2+2+2.1) {$0.75$};
        \node (B) at (-0.95-0.4,-0.1+2+1.05*3) {$1$};

        \node[sloped] (B) at (-1.6-0.4,-0.5+2+1.05) {\rotatebox{90}{Integrated histogram (ECDF)}};
        % \node (B) at (-1.,5.7) {\textbf{b}};
    
    \end{tikzpicture}
    \caption{\textbf{Integrated histograms of gate error}. Integrated histograms for gate errors of qubits, parrallel SG (single-qubit gate) , isolated CZ (CZ gate) and parallel CZ. Their median (dashed lines) are 0.0008 , 0.0095 and 0.016 respectively. 
    }
    \label{SgCz}
\end{figure}

After performing gate calibrations, we use randomized benchmarking (RB) to evaluate the fidelity of single-qubit gates (SG) and CZ gates. As shown in Fig.~\ref{SgCz}, the median error rate for parallel single-qubit gates is below $10^{-3}$, while for parallel CZ gates it is below $2 \times 10^{-2}$. The isolated CZ gate (blue line) achieves an error rate of less than $10^{-2}$.

To estimate the average leakage rate per cycle, we measure the $\ket{2}$ population of the qubit as a function of the number of cycles and fit the results to a rate equation, following the approach in Ref.~\cite{googlerep}. The median leakage error rates are 0.0017 for parallel CZ+SG operations (blue) and 0.0002 for parallel SG operations (red).

It is noted that the fidelity of performing CZ gates in parallel is significantly lower than performing two-qubit gates independently. This is primarily caused by crosstalk, parasitic coupling, and the activation of unwanted couplings during the control process. Crosstalk can couple control signals from certain channels to unexpected channels, causing differences in the modulation signals experienced by the target qubits and the coupler when performing simultaneous CZ gates compared to those experienced during isolated calibration. Parasitic coupling and unwanted coupling lead to the spectator effect, where the states of surrounding qubits influence the evolution of the target qubit pair. These affect the gate fidelity. At the same time, these effects can also cause correlated errors in the system, as shown in Fig.~\ref{fig:pij}\textbf{b}.

\begin{figure}[H]
    \centering
    \begin{tikzpicture}
        \node[anchor=south west,inner sep=0] (image) at (-2,-0.5) {\includegraphics[width=0.5\textwidth]{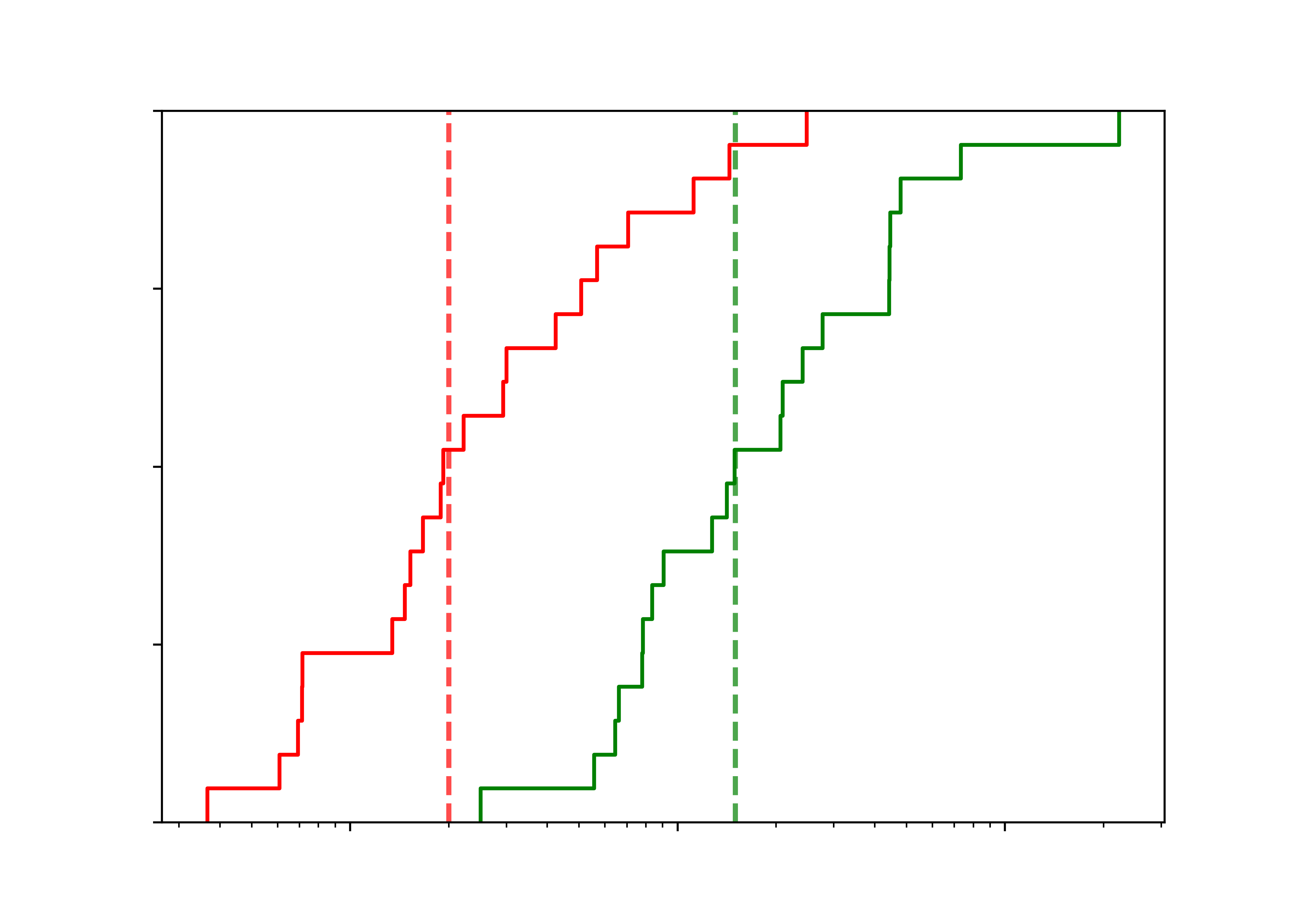}};
        \draw[blue] (-0.8,4.75)--(-0.2,4.75);
        \node[color=blue] (B) at (1.1,4.8) {parallel CZ+SG};

        \draw[red] (-0.8,4.35)--(-0.2,4.35);
        \node[color=red] (B) at (0.7,4.4) {parallel SG};

        % \node (B) at (-0.05,-0.1) {$0.0$};
        \node (B) at (0.5,-0.1) {$10^{-4}$};
        \node (B) at (2.9,-0.1) {$10^{-3}$};
        \node (B) at (5.2,-0.1) {$10^{-2}$};
        \node (B) at (2.6,-0.6) {Leakage rate};

        \node (B) at (-0.95-0.4,-0.65+0.95) {$0$};
        \node (B) at (-0.95-0.4,-0.55+2.) {$0.25$};
        \node (B) at (-0.95-0.4,-0.35+2+1.) {$0.5$};
        \node (B) at (-0.95-0.4,-0.2+2+2.1) {$0.75$};
        \node (B) at (-0.95-0.4,-0.1+2+1.05*3) {$1$};

        \node[sloped] (B) at (-1.6-0.4,-0.5+2+1.05) {\rotatebox{90}{Integrated histogram (ECDF)}};
        % \node (B) at (-1.,5.7) {\textbf{b}};
    
    \end{tikzpicture}
    \caption{\textbf{Integrated histograms of leakage error}. Integrated histograms for leakage errors of qubits, parallel CZ+SG(blue) and parallel SG(red). Their median (dashed lines) are 0.0017 and 0.0002, respectively. 
    }
    \label{Lrate}
\end{figure}

\begin{figure}[htbp]
    \centering
    \includegraphics[width=1\linewidth]{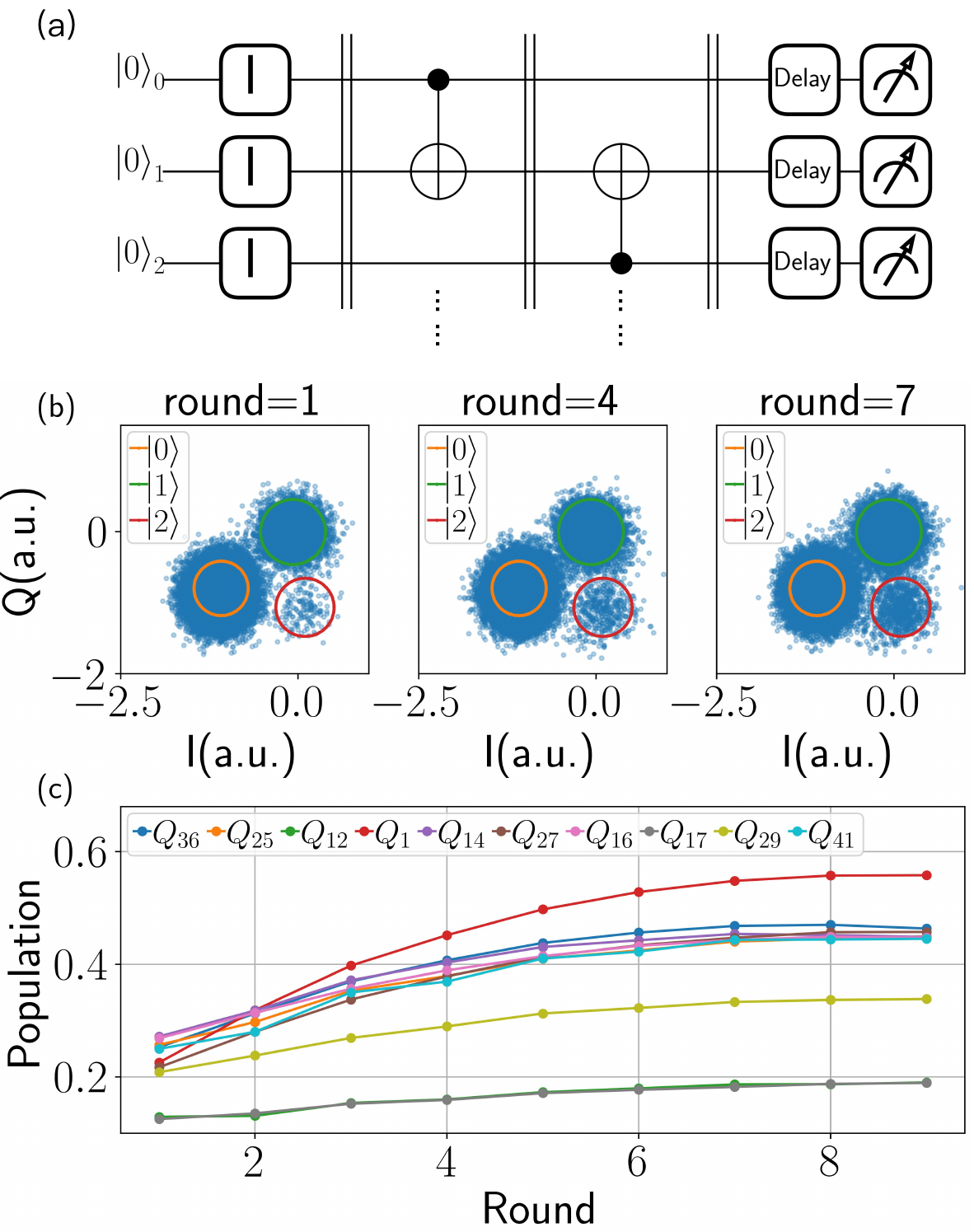}
    \caption{\textbf{Leakage analysis}. (a) The circuit unit for characterizing leakage errors in the repetition code experiment, with repeated measurements conducted in multiple rounds. Even-numbered qubits represent data qubits, odd-numbered qubits represent measurement qubits, and ellipses indicate the extension of the circuit structure for additional qubits.
(b) The in-phase and quadrature components of specific qubit's readout signals during the execution of the circuit in (a). The orange, green, and red circles represent the positions of the readout signals corresponding to qubits in states 0, 1, and 2, respectively. As the number of measurement rounds increases, the proportion of state-2 increases significantly, indicating that more frequent measurements lead to greater leakage of quantum information out of the computational subspace.
(c) The average detection events for different qubits as a function of the number of measurement rounds, distinguishing only between readout signals of states 0 and 1. Higher average values exceeding 0.5 indicate more severe state leakage errors. It is evident that Q1 experiences the most significant state leakage errors.}
    \label{fig:read_leakage}
\end{figure}

\subsection{$p_{ij}$ analysis for configuring decoding priors}

To enhance the accuracy of the decoding process, it is crucial to provide the decoder with a set of prior probability distributions that closely reflect the actual noise conditions of the hardware. These probabilities correspond to the error mechanisms in the decoder error model (DEM) and form the weights of the edges in the decoding graph. This set of prior values is derived through a method known as "$P_{ij}$ analysis." This method aims to calculate the correlation between specific detector configurations by statistically analyzing the frequency of their simultaneous occurrences across numerous samples, using this correlation to characterize probability. Further details can be found in~\cite{googlerep}.

\begin{figure}[htbp]
    \centering
    \includegraphics[width=\linewidth]{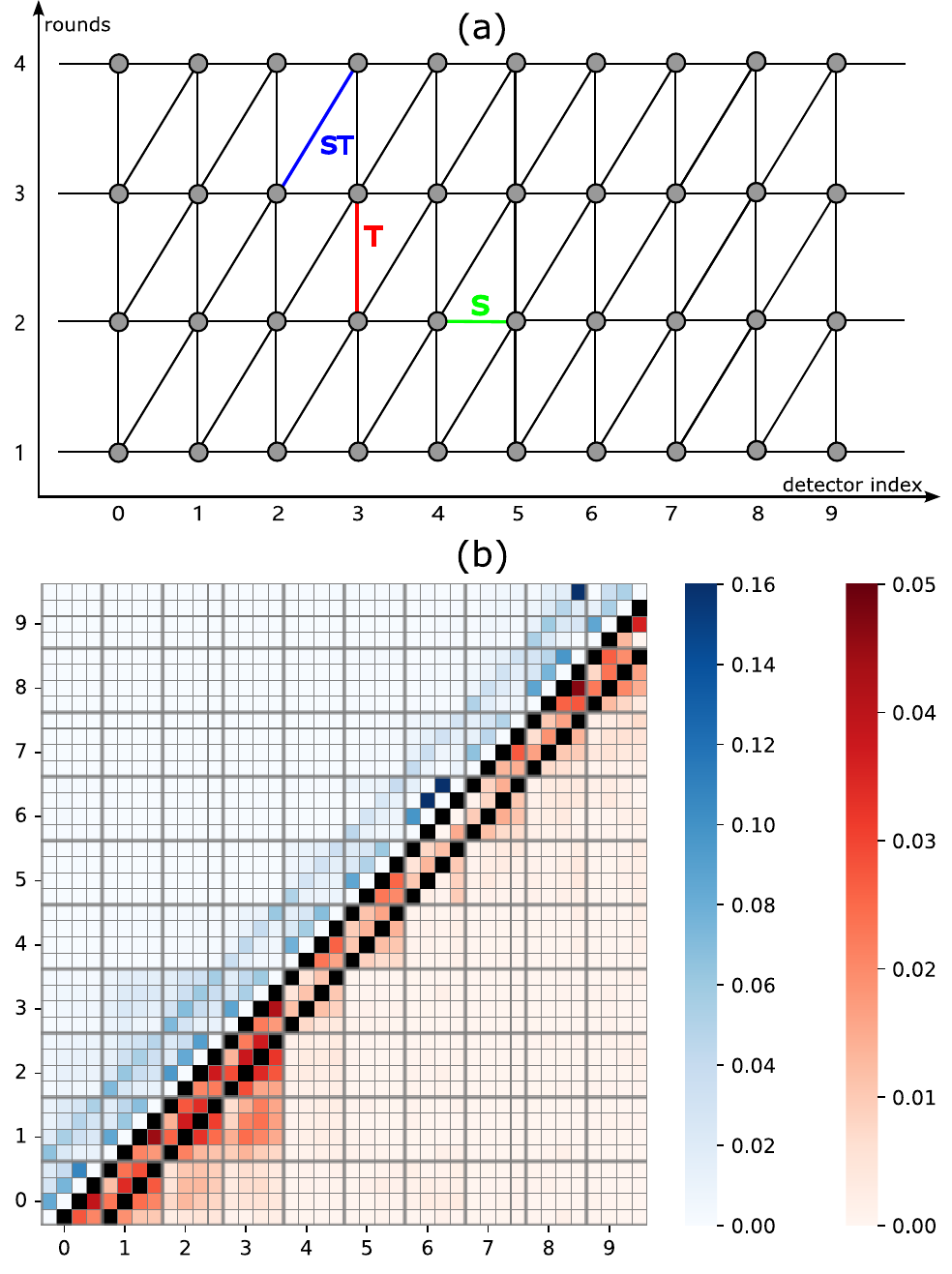}
    \caption{(a) The Detector Error Model (DEM) for a distance-11 round-3 repetition code is depicted, featuring four rounds of detectors. The final round of detectors is derived from the final measurement of all data qubits. Different types of edges are represented in various colors. (b) The heatmap of the $P_{ij}$ correlation analysis is shown. The large blocks along the x and y axes, divided by thicker gray lines, correspond to the detector indices in (a). Within each large block, smaller divisions marked by thin gray lines represent different rounds, corresponding to the y-axis in (a)}
    \label{fig:pij}
\end{figure}

\begin{figure*}[htbp]
    \centering
    \includegraphics[width=\linewidth]{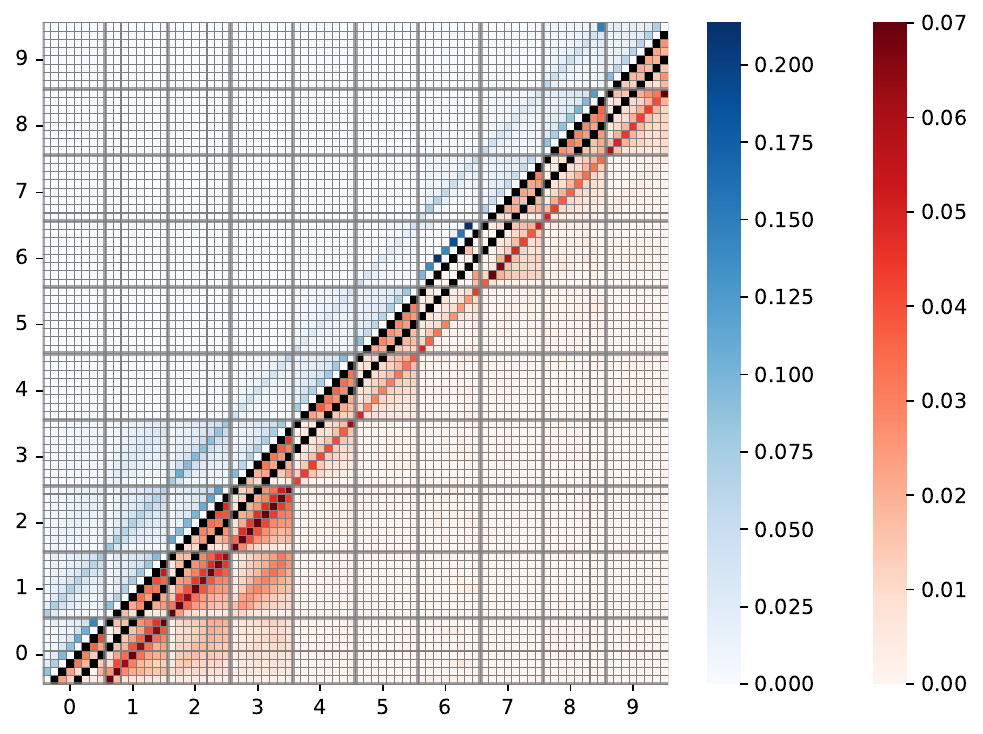}
    \caption{The heatmap of the $P_{ij}$ correlation analysis for distance $11$, rounds $7$. The presence of spacetime (ST) error edges are clearly visible}
    \label{fig:pij_highD}
\end{figure*}

Fig.~\ref{fig:pij}(b) presents the results of the $P_{ij}$ correlation analysis for a distance-11 round-3 repetition code, with its corresponding decoding graph shown above. The x and y axes are divided by thicker gray lines to indicate different detectors, while the smaller divisions within these blocks, marked by thin gray lines, represent different rounds of the same detector. The heatmap is symmetric along the line $y = x$, as the correlation between $i$ and $j$ is identical to that between $j$ and $i$. To better visualize these correlations, we employ two color bars to distinguish between different value scales: the upper-left section of the heatmap uses a blue color bar for higher correlations, while the lower-right section uses a red bar for relatively lower correlations. The black blocks in the lower-right part are masked to subtract the highest values, which are displayed in the symmetric upper-left location. These masked blocks correspond to error mechanisms that connect adjacent rounds of the same detector and neighboring detectors within the same round (i.e., red T (timelike error) edge and green S (spacelike error) edge in Fig.~\ref{fig:pij}(a)). Notably, the timelike error probability for the $6$-th detector shows an unnaturally high value as the rounds progress, indicating poor performance of the 6th ancilla qubit, which aligns with our evaluation shown in Fig.~\ref{fig:read_leakage}. Additionally, the pronounced red areas spilling over in the lower-left corner indices $0$, $1$, $2$, and $3$ of the detector can be attributed to the effects caused by the diffusion of leakage and crosstalk. At the end of the article, we have included a heatmap Fig.~\ref{fig:pij_highD} of the $P_{ij}$ correlation analysis for a distance-11 round-7 repetition code. In this heatmap, the presence of spacetime error (ST) edges is clearly visible.

However, this method relies on a strong assumption: it presumes that correlations exist only between selected specific detectors. This assumption holds in an ideal circuit-level error model, where it is assumed that only Pauli errors are present. Under this assumption, each Pauli error causes the simultaneous flipping of at most two adjacent detectors, meaning that each Pauli error that fits this assumption can be incorporated into a single edge (error mechanism), as illustrated in Fig.~\ref{fig:simple_dem}. Nevertheless, real quantum hardware includes a small number of error mechanisms beyond the error model, which can lead to simultaneous flipping of multiple neighboring detectors or long-range correlations between detectors. These additional error types not only interfere with the calculation of probabilities for error mechanisms within the model but also are not currently accounted for by the decoder, thereby reducing decoding performance.

Another issue arising from these errors beyond the standard circuit-level error model is that they can sometimes lead to deviated or even non-physical values when calculating the probabilities of error mechanisms at the boundary (i.e., edges linked to only one detector). This problem occurs because the method for calculating probabilities at the boundary involves subtracting the probabilities of all other edges connected to the same detector from the total probability of that detector flipping. The additional multi-body correlations caused by non-Pauli errors increase the sum of the probabilities to be subtracted, potentially resulting in deviated or negative probabilities for boundary edges.

Part of this issue can be attributed to the non-reset repetition code we use, which cannot effectively suppress leakage diffusion. When the number of distance and rounds are limited, the impact of beyond-Pauli errors on performance is not so significant, only some deviations occur. However, when the distance and rounds are equal to or greater than $7$, negative probability issues may sometimes arise, necessitating correction. We simply adjust those possible deviation and the one to three non-physical boundary probability values to an empirical value (equal to a certain empirical constant), which still yields physical results after correction. However, when the number of distance and rounds is greater than or equal to 11, the corrected decoding results do not demonstrate a decreasing trend in the logical error rate with increasing distance.

\FloatBarrier

\end{document}